\documentclass[]{interact}
\usepackage{longtable}
\usepackage{tabularx}
\usepackage{subcaption}	
\usepackage{epstopdf}

\usepackage{caption}
\usepackage{csquotes}
\usepackage[numbers,sort&compress]{natbib}
\bibpunct[, ]{[}{]}{,}{n}{,}{,}
\makeatletter
\def\NAT@def@citea{\def\@citea{\NAT@separator}}
\makeatother

\theoremstyle{plain}

\theoremstyle{definition}

\theoremstyle{remark}

\def\bm1{\mbox{\boldmath $1$}}

\def\Ripi1{\mbox{$R_{i1, ii'}$}}

\def\Riip1{\mbox{$R_{i'1, ii'}$}}

\def\X1ii{\mbox{ $X_{1,ii'}$}}

\def\x0ii{\mbox{$\boldmath{x}_{0,ii'}$}} 

\def\H1ii{\mbox{ $H_{1,ii'}$}}
\def\x1ij{\mbox{$x_{1,ij}$}}

\begin{document}


\title{Monte Carlo Expectation-Maximization algorithm to detect imprinting and maternal effects for discordant sib-pair data}

\author{
\name{Ruwani Herath \textsuperscript{a} and Alex Trindade\textsuperscript{b} and Fangyuan Zhang\textsuperscript{c}\thanks{Corresponding author: Fangyuan Zhang, email: fangyuan.zhang@ttu.edu}}
\affil{\textsuperscript{a,b,c}Department of Mathematics and Statistics, Texas Tech University, Lubbock, TX 79409-1042, USA}
}

\maketitle

\begin{abstract}
Numerous statistical methods have been developed to explore genomic imprinting and maternal effects, which are causes of parent-of-origin patterns in complex human diseases. Most of the methods, however, either only model one of these two confounded epigenetic effects, or make strong yet unrealistic assumptions about the population to avoid over- parameterization. A recent partial likelihood method ($LIME_{DSP}$) can identify both epigenetic effects based on discordant sibpair family data without those assumptions. Theoretical and empirical studies have shown its validity and robustness. As $LIME_{DSP}$ method obtains parameter estimation by maximizing partial likelihood, it is interesting to compare its efficiency with full likelihood maximizer. To overcome the difficulty in over-parameterization when using full likelihood, this study proposes a discordant sib-pair design based Monte Carlo Expectation Maximization ($MCEM_{DSP}$) method to detect imprinting and maternal effects jointly. Those unknown mating type probabilities, the nuisance parameters, are considered as latent variables in EM algorithm. Monte Carlo samples are used to numerically approximate the expectation function that cannot be solved algebraically. Our simulation results show that though this $MCEM_{DSP}$ algorithm takes longer computation time, it can generally detect both epigenetic effects with higher power, which demonstrates that it can be a good complement of $LIME_{DSP}$ method.
\end{abstract}



\section{Introduction}
A genome-wide association study (GWAS) is an observational study of a genome-wide set of genetic variants in different individuals to see if any variant is associated with a trait. When applied to human data, GWAS compares participants' DNA with varying phenotypes for a particular trait or disease. GWAS is a powerful tool for identifying hundreds of genetic variants associated with complex human diseases and traits and has provided valuable insights into their genetic architecture. However, most of the variants identified have explained only a small proportion of complex diseases or traits, leading many to question \enquote{missing heritability}\cite{1}, which refers to the fact that single genetic variations cannot account for much of the heritability of diseases, behaviors, and other phenotypes. Other mechanisms may be also involved in this process, such as epigenetic modifications and transcriptional/ translational regulation\cite{3}\cite{4}. In biology, epigenetics studies the heritable phenotype changes that do not involve alterations in the DNA sequence\cite{5}. Thus, the epigenetic factors, including imprinting and maternal genotype effects, have become a research focus to detect missing heritability \cite{6}.\\ 
	Genomic imprinting is an effect of an epigenetic process involving methylation and histone modifications that can silence the expression of a gene inherited from a particular parent without altering the genetic sequence. In diploid organisms, the somatic cells possess two copies of the genome, one inherited from the father and one from the mother. Each autosomal gene is therefore represented by two copies of alleles, with one copy inherited from each parent at fertilization. In mammals, however, a small proportion (\textless 1\%) of genes are imprinted, which means gene expression occurs from only one allele partially or completely \cite{9}. The expressed allele is dependent upon its parental origin, i.e., unequal expression of a heterozygous genotype depending on whether the imprinted variant is inherited from the mother (maternal imprinting) or from the father (paternal imprinting). The imprinting effect is considered as a critical factor in understanding the interplay between the epigenome and genome \cite{10}. For many diploid genes, even if the copy inherited from one parent is defective, there is a substitute allele from the other parent. However, in the case of imprinting, even though there are two copies of the gene, it behaves as a haploid (having a single set of unpaired chromosomes) for this gene because only one copy is expressed. In other words, no substitute allele makes imprinted genes more vulnerable to the adverse effects of mutations. Additionally, genes and mutations that might be generally recessive can be expressed if a gene is imprinted and the dominant allele is silenced \cite{12}. Therefore, the disease can occur due to deletions or mutations in imprinting genes.\cite{13}. Most common conditions involving imprinting include Beckwith-Wiedemann syndrome, Silver-Russell syndrome, pseudohypoparathyroidism \cite{14}, and transient neonatal diabetes mellitus can also involve imprinting \cite{15}.\\
	Maternal effect is the situation where the phenotype of an organism is determined not only by the environment it experiences and its genotype, but also by the environment and genotype of its mother. In genetics, the maternal effects occur when the genotype of its mother determines the phenotype of an organism, irrespective of its own genotype \cite{11}. This effect often occurs because the mother supplies mRNA or proteins during pregnancy. Maternal effect can be considered as a significant reason for varieties of diseases such as those that are related to pregnancy outcomes, childhood cancers and congenital disabilities \cite{16}, certain psychiatric illnesses \cite{17}, and some pregnancy complications \cite{18}.  \\
	Parent-of-origin effects occur when the phenotypic effect of an allele depends on whether it is inherited from an individual’s mother or father \cite{19}. Since both imprinting and maternal effects have parent-of-origin patterns \cite{7} \cite{8}, they can be confounded with each other. For example, paternal imprinting can mimic  maternal effect. Therefore, to identify imprinting and maternal effect, family-based data are needed to find the inheritance path, and it is better to detect the two epigenetic effects jointly.\\
	
	\subsection{Existing methods for detecting Imprinting and Maternal effects}
	Both parametric and non-parametric methods are suggested to study the imprinting and maternal effects. A Parental-Asymmetry Test (PAT) with case-parents trio data was proposed to detect the imprinting effect in nuclear families in the absence of maternal effect\cite{41}\cite{42}. It was extended to accommodate general pedigree data by proposing Pedigree Parental-Asymmetry Test (PPAT), which uses all informative family trios. In addition, to accommodate pedigree data with missing genotypes, a Monte Carlo sampling based PPAT (MCPPAT) method was further developed\cite{43}. The non-parametric methods are mainly used for detecting imprinting effect while assuming there are no maternal effect\cite{32}. Though these non-parametric methods may be more powerful if there is indeed no maternal effect, under the existence of maternal effects, these methods suffer from potential confounding between imprinting and maternal effects, that can inflate false positive or false negative rates\cite{33}. \\
	Numerous parametric methods can identify imprinting and maternal effects jointly based on full likelihood\cite{21}. Case-parent triads and case-mother pairs are two popular study designs\cite{20}. Almost all these methods, however, depend on strong yet unrealistic assumptions concerning mating type probabilities (nuisance parameters) to avoid over-parameterization, with the Logarithm Likelihood Ratio test (LL-LRT) as a classic example\cite{22}. A partial Likelihood method for detecting Imprinting and Maternal effects (LIME) was then proposed as an exception. LIME can avoid over-parameterization by deriving a partial likelihood that is free of the nuisance parameters through matching the case families with control families of the same familial genotypes\cite{20}\cite{23}. LIME is powerful and robust, but it requires the recruitment of separate control families which can be hard to obtain. Therefore, a LIME method based on a Discordant Sib-Pair design $LIME_{DSP}$ was proposed to receive the benefit of LIME without the requirement of separate control families.  Discordant Sibpair design refers to that a nuclear family is recruited if there are at least an affected sibling and an unaffected sibling \cite{25}. Similarly as LIME method, $LIME_{DSP}$ derives partial likelihood by matching affected proband-parent triads with unaffected proband-parent triads and factoring out common terms involving mating type probabilities. \\
	Theoretical and empirical studies have shown the validity and robustness of LIME and $LIME_{DSP}$. The two methods, however, might lack of efficiency in parameter estimation, as they estimate parameters by maximizing partial likelihood, rather than full likelihood. Therefore, it is interesting to compare the estimation efficiency of these partial likelihood methods with a full likelihood method. To avoid over-parameterization problem in full likelihood, a Monte Carlo EM algorithm was proposed earlier for case - control family data. The results showed that the MCEM algorithm can detect both epigenetic effects with higher power and smaller standard error compared with the LIME method \cite{51}. In this study, we will propose a discordant sibpair design based MCEM algorithm ($MCEM_{DSP}$) to detect both epigenetic effects and compare its performance with $LIME_{DSP}$.
	
	\subsection{MCEM algorithm}
	When missing values or latent variables exist, the direct computation of the Maximum Likelihood Estimate (MLE) is not usually achievable. Instead, we use expectation maximization algorithm, which can deal with models that depend on unobservable latent variables. The EM algorithm considers the fact that the distribution of the complete data, which include both observed data and latent variables, can be simpler to deal with \cite{45}. Each EM iteration consists of two steps as the Expectation (E) and the Maximization (M) steps. The E - step computes the expected log-likelihood density of the complete data, conditionally on the observed data and the current parameter value. The parameter is updated in the M - step by maximizing the conditional expectation. The Monte Carlo EM (MCEM) algorithm is a modification of the EM algorithm which is used when the E - step in EM algorithm is not available in closed form. In MCEM, Monte Carlo simulations are used to generate realizations of the conditional hidden data through Markov Chain Monte Carlo (MCMC) routines such as the Gibbs and Metropolis - Hasting sampler \cite{44}. Then the expectation in the E - step is replaced by the empirical mean of the complete log-likelihood. This is the principle of MCEM algorithm proposed by Wei and Tanner (1990) \cite{46}. In our study to detect imprinting effect and maternal effect, as we have nine unobservable mating type probabilities involved in the full likelihood, we will overcome the over-parameterization problem by using the nuisance parameters as latent variables in MCEM algorithm to get maximum likelihood estimation. \\
	In this dissertation, we seek to propose Monte Carlo Expectation Maximization algorithm ($MCEM_{DSP}$) to detect imprinting and maternal effect for discordant sibpair data. First, we have applied $MCEM_{DSP}$ algorithm for discordant sibpair data we generated and compared the results with $LIME_{DSP}$, which is an existing method for detecting imprinting and maternal effects based on partial likelihood. A major difficulty of using $MCEM_{DSP}$ is the high computation time. Secondly, we propose importance sampling based $MCEM_{DSP}$ as solution for the higher computation time of $MCEM_{DSP}$. It can give the similar results as $MCEM_{DSP}$ in shorter time. Finally, we applied $MCEM_{DSP}$ method to Framingham Heart Study (FHS) data to illustrate the utility of $MCEM_{DSP}$. \\
	Chapter 02 contains general introduction for EM and MCEM algorithm with a Gaussian mixture data as an application example. Then we demonstrate the over-parameterization problem in the full likelihood for detecting the epigenetic effects based on discordant sibpair data, which is the reason for the assumptions concerning mating type probabilities in most existing methods. Then we propose the $MCEM_{DSP}$ method to overcome the over-pamameterization problem by using mating type probabilities as the latent variables. To increase the time efficiency, we further develop an importance sampling based $MCEM_{DSP}$ method.\\ 
	Chapter 03 contains extensive simulations of discordant family data with and without additional siblings for eight disease models and eight scenarios. We apply $MCEM_{DSP}$ and $LIME_{DSP}$ to the simulated data, and compare the type I error, power, and relative bias for parameter estimation. In addition, we also use the simulation data to demonstrate the time efficiency for importance sampling based $MCEM_{DSP}$. \\
	Chapter 04 contains real data analysis. We apply $LIME_{DSP}$ and $MCEM_{DSP}$ to Framingham Heart Study data to identify SNPs with association, imprinting or maternal effects related to hypertension. \\

\section{MCEM algorithm}
\subsection{Expectation Maximization algorithm}
	In this section, we summarize the expectation maximization algorithm which is known as the EM algorithm. This is a method to get likelihood maximizer when there are missing data or latent variables.\\
	Let $Y$ be the random vector corresponding to the observed data $y$ with p.d.f. postulated as $g(y,\boldsymbol{\psi})$, where $\boldsymbol{\psi}$ = $\big(\psi_1, ...., \psi_d\big)^d$ is a vector of unknown parameter with parameter space $\Omega$. In the context of EM algorithm, we consider the observed data y as incomplete and the complete data may contain some variables that are never observable. We define $x$ as complete data and we let $z$ denote the vector containing the additional data, referred to as the latent variables or missing data. Consider the EM algorithm as follows:\\
	
	\begin{enumerate}
		\item Let $g_c(x,\boldsymbol{\psi})$ denote the p.d.f. of the random vector $X$ corresponding to the complete data vector $X = (Y,Z)$, where $Y$ is the observed data, and $Z$ is missing data or latent variable. Then we have the following complete-data log likelihood function as:\\
		\begin{equation}
			log L_c(\boldsymbol{\psi}) = g_c(x,\boldsymbol{\psi})
		\end{equation} 
		
		\item \textbf{E-Step:} Let $\boldsymbol{\psi}^{(k)}$ be values for $\boldsymbol{\psi}$ at the iteration $k$. Then, the E-step requires the calculation of:\\
		\begin{equation}
			Q\big(\boldsymbol{\psi},\boldsymbol{\psi}^{(k)}\big) = E_{\boldsymbol{\psi}^{(k)}}[log L_c(\boldsymbol{\psi})|Y]
		\end{equation}
		
		Above equation simply says that at iteration $k$, we need to calculate the expectation of complete likelihood given the information we have for the observable data $y$ and values $\boldsymbol{\psi}^{(k)}$ for the unknown parameter $\boldsymbol{\psi}$.\\

		\vspace{0.5cm}\item \textbf{M-Step:} In M-step, we calculate the maximization of $Q\big(\boldsymbol{\psi},\boldsymbol{\psi}^{(k)}\big)$ with respect to $\boldsymbol{\psi}$ over the parameter space $\Omega$ and denote it as $\boldsymbol{\psi}^{(k+1)}$. It can be shown that,\\
		\begin{equation}
			Q\big(\boldsymbol{\psi},\boldsymbol{\psi}^{(k+1)}\big) \geq Q\big(\boldsymbol{\psi},\boldsymbol{\psi}^{(k)}\big)
		\end{equation}
		and 
		\begin{equation}
			L(\boldsymbol{\psi}^{(k+1)},Y) \geq L(\boldsymbol{\psi}^{(k)},Y)
		\end{equation}
		for all $\boldsymbol{\psi}$ $\in$ $\Omega$.\\
		We continue with the E-step and M-step until the difference $\boldsymbol{\psi}^{(k+1)} - \boldsymbol{\psi}^{(k)}$ changes by an arbitrary small amount.\\
		In practice, a EM sequence will converge to a compact connected set of local maxima of the likelihood function; this limit set may or may not consist of a single point \cite{28}.
	\end{enumerate} 
\subsection{Monte Carlo Expectation Maximization algorithm}
	As in the previous section, in E-step of EM algorithm we need to take the expectation with respect to the conditional distribution of the missing data $Z$ given the observed data $Y = y$. In many cases, however, evaluating the expectation in E-step of the EM algorithm does not yield closed form of the solutions.To overcome this problem, we can simulate random draws from target conditional distribution $Z$ and approximate Q - function by Monte Carlo integration. To proceed this, let $z_1,....,z_m$ be random samples from $f(z|y,\boldsymbol{\psi}^{(k)})$. Then we can use a Monte Carlo approximation to the expectation of EM algorithm as follow:
	\begin{equation}
		Q_{MC}(\boldsymbol{\psi}|\boldsymbol{\psi}^{(k)};y) = \frac{1}{m} \sum_{t=1}^{m}\log L_c(\boldsymbol{\psi}|Y,Z^{(t)})
	\end{equation}
	The update $\boldsymbol{\psi}^{(k+1)}$ is the value of $\boldsymbol{\psi}$ that maximizes $Q_{MC}(\boldsymbol{\psi}|\boldsymbol{\psi}^{(k)};y)$.  
	\subsection{Convergence of MCEM algorithm}
	The expectation-maximization algorithm is an algorithm for maximizing likelihood functions, specially when there are missing data or latent variables exist. If EM algorithm converges, it converges to a stationary point of the likelihood function. Even when the EM algorithm converges, there is no guarantee in general that it converges to a global maximum. However, the limit of an EM sequence can be a local maximum \cite{35} or a saddle point \cite{36}. Further, the convergence rate of the EM algorithm cannot be super linear \cite{37}. When EM works well, the output of the MCEM algorithm is a sequence of parameter values that converges to the maximum likelihood estimate (MLE) \cite{34}. For a given suitable initial value, a sequence of parameter values generated by the Monte Carlo Expectation Maximization algorithm will get arbitrarily close to a maximizer of the observed likelihood with a high probability \cite{38}.\\
	\subsection{Metropolis-Hastings Algorithm}
	The Monte Carlo EM algorithm is a stochastic version of the EM algorithm using MCMC methods to approximate the conditional distribution of the hidden data. A general method for generating samples from posterior distribution is Metropolis-Hastings algorithm.\\
	
	\subsubsection*{\textbf{Algorithm}:}
	\begin{enumerate}
		\item Sample a candidate value $Z^*$ from a proposal distribution $g(.|z^{(t)})$\\
		\item Compute Metropolis-Hastings ratio $R(z^{(t)},Z^*)$\\
		\begin{equation}
			R(u,v) = \frac{f(v)g(u|v)}{f(u)g(v|u)},
		\end{equation}
		where $f(.)$ is the probability density function of the stationary distribution. 
		\item Sample a value for $Z^{(t+1)}$ according to the following:\\
		\begin{equation}
			Z^{(t+1)} =
			\begin{cases}
				Z^* & \textrm{ with probability } {\min [R(Z^{(t)},Z^{\*})]}\\
				Z^{(t)} & \textrm{ otherwise}\\ 
			\end{cases}
		\end{equation}
		
		\item Increment t and return to step 1 until it reaches to the stopping rule.\\ 	
	\end{enumerate}

\subsection{Importance Sampling Method}
	We'll consider the situation where the sample of the latent variables $Z_1,...Z_m$ in the E-step is obtained from a MCMC routine such as Gibbs sampler or Metropolis-Hasting algorithm with stationary distribution $g(Z|Y,\boldsymbol{\psi})$. Drawing an MCMC sample in each iteration of the MCEM algorithm could be very time consuming. We can use importance sampling to improve the computational expenses of the MCEM algorithm \cite{44}. \\
	We initialize the algorithm with a sample $Z_1, Z_2,..., Z_m$ from the distribution $g(Z|Y,\boldsymbol{\psi}^{(0)})$, where $\boldsymbol{\psi}^{(0)}$ is the initial value of the parameter $\boldsymbol{\psi}$ at the start of EM algorithm. \\
	Then, at each iteration r, rather than obtaining a new sample from $g(Z|Y,\hat{\boldsymbol{\psi}}^{(r)})$, with the most recent iterate, we can importance weight the original sample through the updated distribution $g(u|y,\boldsymbol{\psi}^{(r)})$That is,\\
	\begin{equation}
		Q_m\big(\boldsymbol{\psi} | \hat{\boldsymbol{\psi}}^{(r)} \big) = \frac{\sum_{t=1}^{m} w_t \ln f(Y,Z_t | \boldsymbol{\psi})}{\sum_{t=1}^{m}w_t}
	\end{equation}
	
	where\\
	\begin{equation}
		w_t = \frac{g(Z_t|Y,\hat{\boldsymbol{\psi}}^{(r)})}{g(Z_t|Y,\hat{\boldsymbol{\psi}}^{(0)})}
	\end{equation}
	
	In addition, we can show that
	\begin{equation}
		w_t = \frac{L(\hat{\boldsymbol{\psi}}^{(r)}|Z_t,Y) / L(\hat{\boldsymbol{\psi}}^{(r)}|Y)}{L(\hat{\boldsymbol{\psi}}^{(0)}|Z_t,Y) / L(\hat{\boldsymbol{\psi}}^{(0)}|Y)}
	\end{equation}
	
	The likelihood $L(\boldsymbol{\psi}|y)$ can be canceled in formula $(2.4.3)$ and,\\
	\begin{equation}
		Q_m(\boldsymbol{\psi} | \hat{\boldsymbol{\psi}}^{(r)}) = \frac{\sum_{t=1}^{m} w_t^{'} \ln f(Y,Z_t | \boldsymbol{\psi})}{\sum_{t=1}^{m}w_t^{'}}
	\end{equation}
	
	where,
	\begin{equation}
		w_t^{'} = \frac{L(\hat{\boldsymbol{\psi}}^{(r)} | Z_t,Y)}{L(\hat{\boldsymbol{\psi}}^{(0)} | Z_t,Y)}
	\end{equation}
	
	This choice will not affect the EM algorithm because the unknown normalizing constant\\
	$$\frac{L(\hat{\boldsymbol{\psi}}^{(0)}|Y)}{L(\hat{\boldsymbol{\psi}}^{(r)}|Y)}$$ 
	does not depend on the unknown value of $\boldsymbol{\psi}$, and it does not come into play in the maximization step. \\
	
	\subsubsection*{\textbf{Algorithm:}}
	\begin{enumerate}
		\item Initialize $\boldsymbol{\psi}^{(0)}$.
		\item  Generate $Z_1, ....Z_m \sim g(Z | Y ,\boldsymbol{\psi}^{(o)})$ via an MCMC algorithm.\\
		\item  At iteration $r+1$ compute the importance weights\\
		$$w_t^{(r)} = \frac{L(\hat{\boldsymbol{\psi}}^{(r)} | Z_t,Y)}{L(\hat{\boldsymbol{\psi}}^{(0)} | Z_t,Y)}$$
		\item  \textit{E - step:} Estimate $Q(\boldsymbol{\psi} | \hat{\boldsymbol{\psi}}^{(r)})$ by\\
		$$Q_m(\boldsymbol{\psi} | \hat{\boldsymbol{\psi}}^{(r)}) = \frac{\sum_{t=1}^{m}w_t^{(r)} \ln f(Z_t,Y | \boldsymbol{\psi})}{\sum_{t=1}^{m}w_t^{(r)}}$$
		\item  \textit{M - step:} Maximize $Q_m(\boldsymbol{\psi} | \hat{\boldsymbol{\psi}}^{(r)})$ to obtain $\hat{\boldsymbol{\psi}}^{(r+1)}$ 
		\item Repeat Steps 3 through 5 until convergence.	
	\end{enumerate}
	
	There can be some drawbacks for importance sampling estimators. If the importance density $g(Z_t | Y,\hat{\boldsymbol{\psi}}^{(r)})$ is not close enough to $g(Z_t | Y,\hat{\boldsymbol{\psi}}^{(0)})$, then the weights will vary widely giving many samples little weight and allowing for a few variates to be overinfluetial. Also, if the initial values $\boldsymbol{\psi}^{(0)}$ are poor, the estimator $Q_m\big(\boldsymbol{\psi} | \hat{\boldsymbol{\psi}}^{(r)}\big)$ will take a long time to converge. We can alleviate this problem by initiating burn-in whereby for the first few iterations and a new sample is obtained from $g(Z_t | Y,\hat{\boldsymbol{\psi}}^{(r)})$ rather than importance weighting. \\
	\begin{enumerate}
		\item Initialize $\boldsymbol{\psi}^{(0)}$, and run burn-in.
		\begin{enumerate}
			\item Set importance weights $w_t = 1$ for all $t = 1,...,m$.\\
			At iteration b,
			\item Generate $Z_1,...,Z_m \sim g(Z | Y,\boldsymbol{\psi}^{(b)})$ via an MCMC algorithm.
			\item Run E and M steps above with $r=b$.
			\item Repeat Steps 1(b) and 1(c) for B burn-in iterations.
			\item Reinitialize $\boldsymbol{\psi}^{(0)} = \boldsymbol{\psi}^{(B)}$.
		\end{enumerate}
	\end{enumerate}

\section{Applying Expectation Maximization Algorithm to detecting imprinting and Maternal Effects based on discordant sibpair design} 
	\subsection{Notation and Genetic Model}
	Consider a candidate genetic marker with two alleles $M_1$ and $M_2$, which may code for disease susceptibility or epigenetic effect. We define M and F as the number of variant allele $M_1$ carried by the mother and the father in a nuclear family. They can take values 0,1 or 2, corresponding to genotype $M_2M_2$, $M_1M_2$ or $M_1M_1$, respectively. Let $C_i$ be the random variable denoting the number of $M_1$ alleles carried by the child $i$, $i = 1, 2,....,k$. Specifically, $C_1$ and $C_2$ are designated for the affected and unaffected probands, respectively, through which the discordant sib-pair family is recruited, whereas $C_i, i = 3,4,..,k$ are for the additional siblings. The indicator variable $D_i, i = 1,2,...,k$ denotes the disease status of children where 1 denotes being affected and 0 denotes being normal. Thus, for the affected child and unaffected child $D_1 = 1$ and $D_2 = 0$, respectively. \\
	Assume the disease penetrance follows a multiplicative relative risk model
	\begin{equation}
		P(D = 1|M = m, F = f, C = c) = \delta R_1^{I(c = 1)} R_2^{I(c = 2)} R_{im}^{I(c = 1_m)} S_1^{I(m = 1)} S_2^{I(m = 2)}
	\end{equation}
	where $R_1$ and $R_2$ denote the relative risk due to one or two copies of individual's own variant allele, $R_{im}$ denotes relative risk due to imprinting effect, $S_1$ and $S_2$ denote the relative risk due to one or two copies of the mother's variant allele, and $\delta$ denote the disease phenocopy rate. The notation $c = 1_m$ indicates that the child's genotype is $M_1M_2$ and the variant allele is inherited from mother. We are interested in the model parameter estimation, collectively denoted as $\boldsymbol{\theta} = (\delta, R_1, R_2, R_{im}, S_1, S_2)$. Note that, all parameters are positive. Further, $R_{im} > 1, < 1,$ and $= 1$ represent paternal, maternal, and no imprinting effect, respectively. Although there is no restriction placed on maternal effects, $S_1$ and $S_2$, they are typically $\geq 1$, where the equality denotes no maternal effect. \\
	
	\subsection{The situation when $\mu_{ij}$s are fixed unknown parameters}
	$$Y= \{n_{mfc_1c_2c_3...c_k}\} ; k \geq 2$$\\	
	
	Denote $\mu_{mf}$ as the mating type probability of (M,F) = (\textit{m,f}), that is, probability of parental pairs in which the mothers carry m copies and the fathers f of the variant allele. $n_{mfc_1c_2c_3...c_k}$ for $k>2$ represent the number of families with certain familial genotype combinations. \\ 
	Let $\boldsymbol{\mu} = (\mu_{00},\mu_{01},\mu_{02}, \mu_{10}, \mu_{11}, \mu_{12}, \mu_{20}, \mu_{21}, \mu_{22})$ denote the mating type probabilities which are fixed and unknown.
	We can define full likelihood as \\
	\begin{equation}
		\begin{aligned}[b]
			P(Y;\boldsymbol{\theta},& \boldsymbol{\mu}) = \prod_{i=1}^{n} \big[P(M^{(i)}=m,F^{(i)}=f,C_1^{(i)} = c_1,C_2^{(i)}=c_2,C_3^{(i)}=c_3,\\
			&...,C_k^{(i)}=c_k,D_3^{(i)},...D_k^{(i)}|D_1^{(i)}=1,D_2^{(i)}=0)\big] \\
			&= \prod_{i=1}^{n}\Big[P\big(M^{(i)}=m,F^{(i)}=f,C_1^{(i)} = c_1,C_2^{(i)}=c_2)|D_1^{(i)}=1,D_2^{(i)}=0\big)\\ &.P\big(C_3^{(i)} = c_3, C_4^{(i)} = c_4,...,C_k^{(i)} = c_k, D_3^{(i)}, D_4^{(i)},...,D_k^{(i)}|M^{(i)}=m,F^{(i)}=f\big)\Big]  \\
			&= \prod_{i=1}^{n}\Bigg[\bigg(\frac{  P(M^{(i)}=m,F^{(i)}=f,C_1^{(i)}=c_1,C_2^{(i)}=c_2,D_1^{(i)}=1,D_2^{(i)}=0)}{P(D^{(i)}_1=1,D^{(i)}_2=0)}\bigg)\\ &.P\big(C_3^{(i)} = c_3, C_4^{(i)} = c_4,...,C_k^{(i)} = c_k, D_3^{(i)}, D_4^{(i)},...,D_k^{(i)}|M^{(i)}=m,F^{(i)}=f\big)\Bigg]\\
		\end{aligned}
	\end{equation}
	
	Here, we consider $k \geq 2$, since there should be at least one affected child and one unaffected child. The disease status for additional siblings can be arbitrary. When there are no additional siblings, the second term of the equation will equal to 1. \\
	The numerator of the first term on the right hand side of the equation can be expanded as
	\begin{equation}
		\begin{aligned}[b]
			P(M^{(i)}=m & ,F^{(i)}  =f,C_1^{(i)} =c_1,C_2^{(i)}=c_2,D_1^{(i)}=1,D_2^{(i)}=0) \\
			& = P(M=m,F=f).P(C_1=c_1|M=m,F=f)\\ &.P(C_2=c_2|M=m,F=f).P(D_1=1|M=m,F=f,C_1=c_1)\\
			&.P(D_2=0|M=m,F=f,C_2=c_2)\\
			& = \mu_{mf}.P(C_1=c_1|M=m,F=f).P(C_2=c_2|M=m,F=f)\\ 
			&.P(D_1=1|M=m,F=f,C_1=c_1).P(D_2=0|M=m,F=f,C_2=c_2)
		\end{aligned}
	\end{equation}
	
	where\\
	\begin{equation}
		P(D_1=1|M=m,F=f,C_1=c_1) = \delta R_{1}^{I(c_1=1)} R_{2}^{I(c_1=2)} R_{im}^{I(c_1=1_m)} S_{1}^{I(M=1)} S_{2}^{I(M=2)} 
	\end{equation}
	\begin{equation}
		P(D_2=0|M=m,F=f,C_2=c_2) = 1 - \delta R_{1}^{I(c_2=1)} R_{2}^{I(c_2=2)} R_{im}^{I(c_2=1_m)} S_{1}^{I(M=1)} S_{2}^{I(M=2)}
	\end{equation}
	
	\vspace{0.5cm}
	are obtained according to disease penetrance model.\\
	In total, there are 29 possible combinations of genotypes for parents $(M,F)$, affected child $(C_1)$ and unaffected child $(C_2)$. Table 3.2 lists the corresponding joint probability of mother, father, affected child, unaffected child genotypes and proband disease status $P(M = m,F = f,C_1 = c_1,C_2 = c_2,D_1 = 1,D_2 = 0)$. (See Appendix 2) In the expressions in Table 3.2, the $\mu_{mf}$'s $(m = 0,1,2, f = 0,1,2)$ are the mating type probabilities, i.e., $\mu_{mf} = P(M = m, F = f)$. Note that, we do not make any assumptions about the mating symmetry, and therefore, $\mu_{mf}$ is not necessarily equal to $\mu_{fm}$.\\
	\begin{table}[!htb]
		\caption{Joint probability of mother - father - affected child - unaffected child genotype and proband disease status}
		\centering
		\begin{tabular}{c c c c c c}
			\hline
			Type & m & f & $c_1$ & $c_2$ & $P(M = m,F = f,C_1 = c_1,C_2 = c_2,D_1 = 1,D_2 = 0)$\\
			\hline
			1 & 0 & 0 & 0 & 0 & $\mu_{00}\delta(1-\delta)$\\
			2 & 0 & 1 & 0 & 0 & $\mu_{01}\frac{1}{4}\delta(1-\delta)$\\
			3 & 0 & 1 & 1 & 0 & $\mu_{01}\frac{1}{4}\delta r_1(1-\delta)$\\
			4 & 0 & 1 & 0 & 1 & $\mu_{01}\frac{1}{4}\delta(1-\delta r_1)$\\
			5 & 0 & 1 & 1 & 1 & $\mu_{01}\frac{1}{4}\delta r_1(1-\delta r_1)$\\
			6 & 0 & 2 & 1 & 1 & $\mu_{02}\delta r_1(1-\delta r_1)$\\
			7 & 1 & 0 & 0 & 0 & $\mu_{10}\frac{1}{4}\delta s_1 (1-\delta s_1 )$\\
			8 & 1 & 0 & 1 & 0 & $\mu_{10}\frac{1}{4}\delta r_1 s_1 r_{im}(1-\delta s_1 r_{im})$\\ 
			9 & 1 & 0 & 0 & 1 & $\mu_{10}\frac{1}{4}\delta s_1 r_{im}(1-\delta r_1 s_1 r_{im})$\\
			10 & 1 & 0 & 1 & 1 & $\mu_{10}\frac{1}{4}\delta r_1 s_1 r_{im}(1-\delta r_1 s_1 r_{im})$\\
			11 & 1 & 1 & 0 & 0 & $\mu_{11}\frac{1}{16}\delta s_1(1-\delta s_1)$\\
			12 & 1 & 1 & 1 & 0 & $\mu_{11}\frac{1}{16}\delta r_1 s_1(1-\delta s_1)(1 + r_{im})$\\
			13 & 1 & 1 & 0 & 1 & $\mu_{11}\frac{1}{16}\delta s_1 [2 - \delta r_1 s_1(1 + r_{im})]$\\
			14 & 1 & 1 & 1 & 1 & $\mu_{11}\frac{1}{16}\delta r_1 s_1(1-\delta s_1)(1 + r_{im}) [2 - \delta r_1 s_1(1 + r_{im})]$\\
			15 & 1 & 1 & 2 & 0 & $\mu_{11}\frac{1}{16}\delta r_2 s_1(1-\delta s_1)$\\
			16 & 1 & 1 & 0 & 2 & $\mu_{11}\frac{1}{16}\delta s_1 (1-\delta r_2 s_1)$\\
			17 & 1 & 1 & 2 & 2 & $\mu_{11}\frac{1}{16}\delta r_2 s_1(1 - \delta r_2 s_1)$\\
			18 & 1 & 1 & 1 & 2 & $\mu_{11}\frac{1}{16}\delta r_1 s_1 (1 + r_{im}) (1-\delta r_2 s_1)$\\
			19 & 1 & 1 & 2 & 1 & $\mu_{11}\frac{1}{16}\delta r_2 s_1 [2 - \delta r_1 s_1(1 + r_{im})]$\\
			20 & 1 & 2 & 1 & 1 & $\mu_{12}\frac{1}{4}\delta r_1 s_1(1-\delta r_1 s_1)$\\
			21 & 1 & 2 & 1 & 2 & $\mu_{12}\frac{1}{4}\delta r_1 s_1(1-\delta r_2 s_1)$\\
			22 & 1 & 2 & 2 & 1 & $\mu_{12}\frac{1}{4}\delta r_2 s_1(1-\delta r_1 s_1)$\\
			23 & 1 & 2 & 2 & 2 & $\mu_{12}\frac{1}{4}\delta r_2 s_1(1-\delta r_2 s_1)$\\
			24 & 2 & 0 & 1 & 1 & $\mu_{12}\frac{1}{4}\delta r_2 s_1(1-\delta r_2 s_2)$\\
			25 & 2 & 1 & 1 & 1 & $\mu_{20}\delta r_1 s_2 r_{im}(1-\delta r_1 s_2 r_{im})$\\
			26 & 2 & 1 & 2 & 1 & $\mu_{21}\frac{1}{4}\delta r_1 s_2 r_{im}(1-\delta r_1 s_2 r_{im})$\\
			27 & 2 & 1 & 1 & 2 & $\mu_{21}\frac{1}{4}\delta r_2 s_2(1-\delta r_1 s_2 r_{im})$\\
			28 & 2 & 1 & 2 & 2 & $\mu_{21}\frac{1}{4}\delta r_2 s_2(1-\delta r_2 s_2)$\\
			29 & 2 & 2 & 2 & 2 & $\mu_{22}\delta r_2 s_2(1-\delta r_2 s_2)$\\
		\end{tabular}
	\end{table}
	
	The denominator of the first term of the right hand side of the equation and the second term can be expanded as follows.
	\begin{equation}
		\begin{aligned}[b]
			P(D^{(i)}_1=1 & ,D^{(i)}_2=0) \\ & = \sum_{m,f,c_1,c_2}P(M=m,F=f,C_1=c_1,C_2=c_2,D_1=1,D_2=0)
		\end{aligned}
	\end{equation}
	
	and
	\begin{equation}
		\begin{aligned}[]
			P\big(C_3^{(i)} = c_3 & , C_4^{(i)} = c_4,...,C_k^{(i)} = c_k, D_3^{(i)}, D_4^{(i)},...,D_k^{(i)}|M^{(i)}=m,F^{(i)}=f\big)\\
			& = P(C_3 = c_3|M=m,F=f)....P(C_k = c_k|M=m,F=f)\\
			&.P(D_3|M=m,F=f,C_3 = c_3)...P(D_k|M=m,F=f,C_k = c_k)
		\end{aligned}
	\end{equation}

	Therefore, we can see that the full likelihood includes both parameter vector $\boldsymbol{\theta}$ of interest and nuisance parameter vector $\boldsymbol{\mu}$. It is hard to maximize the full likelihood function directly with the over-parameterization problem. This is the reason why many methods need to make assumptions concerning mating type probabilities, such as mating symmetry ,i.e., $\mu_{ij} = \mu_{ji}$, to reduce the number of parameters. To overcome this problem of over-parameterization without making those strong and unrealistic assumptions, we propose $MCEM_{DSP}$ algorithm to get the likelihood maximizer with the nuisance parameters used as latent variables.

	\subsection{Applying $MCEM_{DSP}$}
	Let the random vector X be complete data that consist of the random vectors Y and Z where\\
	\begin{center}
		$Z_{ij}$ = $\mu_{ij}$ \textit{i} = 0,1,2   \textit{j} = 0,1,2 \\
		\begin{center}
			$Y= \{n_{mfc_1c_2c_3...c_k}\} ; k \geq 2$\\
		\end{center}
		
	\end{center}
	
	In this study, we consider the mating type probabilities are latent variables. We use the Expectation Maximization (EM) algorithm to compute maximum likelihood estimates in the problems with latent variables.\\
	We can define $P(Y|Z = \mu) = P(Y;\mu)$ as equation 2.5.2, where the only difference is mating type probabilities are viewed as latent random variables, rather than fixed parameters. \\
	
	We consider the mating type probabilities $\mu_{ij}$'s, are random values that follow Dirichlet distribution $\pi(Z)$. \\ 
	\begin{equation}
		\pi(Z) = \frac{1}{B(\boldsymbol{\alpha})}\prod_{m=0}^{2}\prod_{f=0}^{2}{\mu_{mf}^{\alpha_{mf}}}^{-1} 
	\end{equation}
	Here, $\boldsymbol{\alpha} = \big(\alpha_{00}, \alpha_{01}, \alpha_{02}, \alpha_{10}, \alpha_{11}, \alpha_{12}, \alpha_{20}, \alpha_{21}, \alpha_{22} \big)$ is the concentration parameter.\\
	To apply EM algorithm, the complete log likelihood function can be defined as\\
	\begin{equation}
		\begin{aligned}[b]
			log  L_c(\psi) & =  log\{P(Y|Z)P(Z)\}\\
			& =  log \Biggl\{\Big[\prod_{i=1}^{n} \big(P(M^{(i)}=m,F^{(i)}=f,C_1^{(i)} = c_1,C_2^{(i)}=c_2,C_3^{(i)}=c_3, \\ &...,C_k^{(i)}=c_k,D_3^{(i)},...D_k^{(i)}|D_1^{(i)}=1,D_2^{(i)}=0)\big)\Big].\pi(Z)\Biggr\}\\
			& = \sum_{i=1}^{n} \Biggl\{ log \Bigg[ \frac{  P(M^{(i)}=m,F^{(i)}=f,C_1^{(i)}=c_1,C_2^{(i)}=c_2,D_1^{(i)}=1,D_2^{(i)}=0)}{P(D^{(i)}_1=1,D^{(i)}_2=0)}\Bigg] \\ 
			& + \log \bigg(P\big(C_3^{(i)} = c_3, C_4^{(i)} = c_4,...,C_k^{(i)} = c_k, D_3^{(i)}, D_4^{(i)}\\ &,...,D_k^{(i)}|M^{(i)}=m,F^{(i)}=f\big)\bigg)\Biggr\} + log  \pi(Z)\\
		\end{aligned}
	\end{equation}
	
	We define $\boldsymbol{\psi}$ as\\
	\begin{center}
		$\boldsymbol{\psi}$ = ($\boldsymbol{\theta}$, $\boldsymbol{\alpha}$) 
	\end{center}
	where $\boldsymbol{\theta}$ = \big($\delta, R_1, R_2, R_{im}, S_1, S_2$\big) and $\boldsymbol{\alpha}$ = \big($\alpha_{00}, \alpha_{01}, \alpha_{02}, \alpha_{10}, \alpha_{11}, \alpha_{12}, \alpha_{20}, \alpha_{21}, \alpha_{22} $\big) \\
	The initial values for $\boldsymbol{\psi}^0 = (\boldsymbol{\theta}^0 , \boldsymbol{\alpha}^0)$ are taken in the way that $\boldsymbol{\theta}^0$ is from the results for $LIME_{DSP}$ method and $\boldsymbol{\alpha^0 = \Bigg[(\frac{n_{mf}}{\sum n_{mf}})\times 100\Bigg] + 1} $.\\
	
	\subsubsection*{\textbf{E-Step}:} 
	For E-step, we calculate $Q(\boldsymbol{\psi};\boldsymbol{\psi}^k)$ as follows\\ 
	\begin{equation}
		\begin{aligned}[b]
			& Q( \boldsymbol{\psi} ;\boldsymbol{\psi}^k) = E_{\boldsymbol{\psi}^k}\{log  L_c(\boldsymbol{\psi})|Y\} \\
			& = E_{\boldsymbol{\psi}^k}\Bigg(\sum_{i=1}^{n} \Bigg\{log \Bigg[\frac{  P(M^{(i)}=m,F^{(i)}=f,C_1^{(i)}=c_1,C_2^{(i)}=c_2,D_1^{(i)}=1,D_2^{(i)}=0)}{P(D_1^{(i)}=1,D_2^{(i)}=0)}\Bigg] \\ 
			& + \log \bigg(P\big(C_3^{(i)} = c_3, C_4^{(i)} = c_4,...,C_k^{(i)} = c_k, D_3^{(i)}, D_4^{(i)},...,D_k^{(i)}|M^{(i)}=m,F^{(i)}=f\big)\bigg)\Bigg\}\\
			& + log  \pi(Z) |Y\Bigg)\\
		\end{aligned} 
	\end{equation}
	
	where\\
	\begin{equation}
		\begin{aligned}[b]
			E_{\boldsymbol{\psi}^k} & \Bigg(\sum_{i=1}^{n} \Bigg\{log \Bigg[\frac{  P(M^{(i)}=m,F^{(i)}=f,C_1^{(i)}=c_1,C_2^{(i)}=c_2,D_1^{(i)}=1,D_2^{(i)}=0)}{P(D_1^{(i)}=1,D_2^{(i)}=0)}\Bigg]\\ 
			& + \log \bigg(P\big(C_3^{(i)} = c_3, C_4^{(i)} = c_4,...,C_k^{(i)} = c_k, D_3^{(i)}, D_4^{(i)}\\ & ,...,D_k^{(i)}|M^{(i)}=m,F^{(i)}=f\big)\bigg)\Bigg\}|Y\Bigg) \\ 
			& = \int\sum_{i=1}^{n} \Bigg\{log \Bigg[\frac{  P(M^{(i)}=m,F^{(i)}=f,C_1^{(i)}=c_1,C_2^{(i)}=c_2,D_1^{(i)}=1,D_2^{(i)}=0)}{P(D_1^{(i)}=1,D_2^{(i)}=0)}\Bigg] \\ 
			& + \log \bigg(P\big(C_3^{(i)} = c_3, C_4^{(i)} = c_4,...,C_k^{(i)} = c_k, D_3^{(i)}, D_4^{(i)}\\ &,...,D_k^{(i)}|M^{(i)}=m,F^{(i)}=f\big)\bigg)\Bigg\}.f_{\psi^k}(Z|Y).dZ
		\end{aligned}
	\end{equation}

	and\\
	\begin{equation}
		\begin{aligned}[b]
			E_{\boldsymbol{\psi}^k}(log  \pi(Z) |Y) & = E_{\boldsymbol{\psi}^k}\Bigg(log\Bigg\{\frac{1}{B(\boldsymbol{\alpha})}\prod_{m=0}^{2}\prod_{f=0}^{2}{Z_{mf}^{\alpha_{mf}}}^{-1}\Bigg\}|Y\Bigg)\\
			& = E_{\boldsymbol{\psi}^k}\Bigg(-logB(\boldsymbol{\alpha}) + (\alpha_{mf} - 1)\sum_{m=0}^{2}\sum_{f=0}^{2}log Z_{mf}\Bigg)\\
			& = \int\Bigg(-logB(\boldsymbol{\alpha}) + (\alpha_{mf} - 1)\sum_{m=0}^{2}\sum_{f=0}^{2}log Z_{mf}\Bigg).f_{\psi^k}(Z|Y).dZ\\
		\end{aligned}
	\end{equation}
	
	Since we cannot calculate the integrals in the above equations explicitly, we use Metropolis-Hastings algorithm to take samples from $f_{\boldsymbol{\psi}^k}(Z|Y)$ and get the Monte Carlo estimation of $Q(\boldsymbol{\psi},\boldsymbol{\psi}^{(k)})$.\\
	Here we use proposal distribution $g(Z) = \pi^{(\boldsymbol{\alpha}^k)}(Z)$. The stationary distribution is $f(Z) = P^{(\boldsymbol{\psi}^k)}(Z|y)$.\\
	Here, 
	\begin{equation}
		\pi^{(\boldsymbol{\alpha}^k)}(Z) = \frac{1}{B(\boldsymbol{\alpha}^k)}\prod_{m=0}^{2}\prod_{f=0}^{2}{Z_{mf}^{\boldsymbol{\alpha}_{mf}^k-1}}
	\end{equation}
	and we consider $\boldsymbol{\alpha}^k$ are obtained from the last EM iteration.\\
	In the first step we take a sample $Z^*$ from $\pi^{(\alpha^k)}(Z)$ distribution. Then we take turns to update two elements of $Z$ random vector at a time to improve the acceptance rate of EM algorithm. Suppose we want to update the first two element of the sample based on the conditional distribution of $Z_{A_1}|Z_{A_2}$ where $Z_{A_1} = (Z_1,Z_2)$ and $Z_{A_2} = (Z_3,...,Z_9)$. Then we can show the conditional distribution is as follows\\
	\begin{equation}
		\begin{aligned}[b]
			f_{Z_1,Z_2|Z_3,...,z_9}(z_1,z_2|z_3,...,z_9) &  
			& = \frac{\Gamma(\sum_{i=1}^{2}\alpha_i)}{\prod_{i=1}^{2}\Gamma(\alpha_i)} \prod_{i=1}^{2} \Bigg[z_i \Bigg(1 - \sum_{j=3}^{k}z_j\Bigg)^{-1}\Bigg]^{\alpha_i - 1} \Bigg(1 - \sum_{j=3}^{k}z_j\Bigg)^{-1}
		\end{aligned}
	\end{equation}
	
	Then, we can calculate Metropolis-Hastings ratio as follow:
	\begin{equation}
		\begin{aligned}[b]
			R(Z^{(t)},Z^*) = \frac{f(Z^*)\pi^{(\boldsymbol{\alpha}^k)}(Z^{(t)})}{f(Z^{(t)})\pi^{(\boldsymbol{\alpha}^{(k)})}(Z^*)} & = \frac{P^{(\boldsymbol{\psi}^k)}(Z^*|Y)\pi^{(\boldsymbol{\alpha}^k)}(Z^{(t)})}{P^{(\boldsymbol{\psi}^k)}(Z^{(t)}|Y)\pi^{(\boldsymbol{\alpha}^k)}(Z^*)}\\ 
			&\propto\frac{P^{(\boldsymbol{\theta}^k)}(Y|Z^*)\pi^{(\boldsymbol{\alpha}^k)}(Z^*)\pi^{(\boldsymbol{\alpha}^k)}(Z^{(t)})}{P^{(\boldsymbol{\theta}^k)}(Y|Z^{(t)})\pi^{(\boldsymbol{\alpha}^k)}(Z^{(t)})\pi^{(\boldsymbol{\alpha}^k)}(Z^*)} = \frac{P^{(\boldsymbol{\theta}^k)}(Y|Z^*)}{P^{(\boldsymbol{\theta}^k)}(Y|Z^{(t)})}\\
		\end{aligned}
	\end{equation}
	
	where $\boldsymbol{\theta}^k$ represents the values for the parameters $\delta, R_1, R_2, R_{im}, S_1, S_2$ from previous EM iteration. \\ 
	Finally, we can find sample $Z^{t+1}$ as follows:\\
	$$Z^{(t+1)} =
	\begin{cases}
		Z^* & \textrm{with probability min}{(R(Z^{(t)},Z^*),1)} \\
		Z^{(t)} & \textrm{otherwise}\\
	\end{cases}	$$ 
	In the $k^{th}$ iteration of the MCEM algorithm, we use MH algorithm to generate 10000 sample points $Z^{(1)}_{kmf}$, $Z^{(2)}_{kmf}$,...,$Z^{(10000)}_{kmf}$ from probability distribution $f_{\boldsymbol{\psi}^{(k)}}\big(Z|Y\big)$ as shown above. 
	We determine 10000 sample points are good enough for MH algorithm to converge, as Brooks Gelman scale reduction factor is very close to 1 for 10000 sample size in simulation  \cite{64}.
	Then we use Monte-Carlo method to estimate integration in E -step as follows
	\begin{equation}
		\begin{aligned}[b]
			&\hat{E}_{\boldsymbol{\psi}^{(k)}}\{log P_{\boldsymbol{\theta}}\big(Y|Z\big)|Y\} \\
			& = \frac{1}{10000}\sum_{t = 1}^{n} \sum_{j=1}^{10000}\bigg[log \bigg(\frac{P(M^{(i)}=m,F^{(i)}=f,C_1^{(i)}=c_1,C_2^{(i)}=c_2,D_1^{(i)}=1,D_2^{(i)}=0)}{P(D^{(i)}_1=1,D^{(i)}_2=0)}\bigg) \\ 
			& + \log \bigg(P\big(C_3^{(i)} = c_3, C_4^{(i)} = c_4,...,C_k^{(i)} = c_k, D_3^{(i)}, D_4^{(i)},...,D_k^{(i)}|M^{(i)}=m,F^{(i)}=f\big)\bigg)\bigg]
		\end{aligned}
	\end{equation}
	
	where \\
	\begin{equation}
		\begin{aligned}[b]
			P(M^{(i)}=m & ,F^{(i)}=f ,C_1^{(i)}=c_1,C_2^{(i)}=c_2,D_1^{(i)}=1,D_2^{(i)}=0)\\
			& = Z^{(j)}_{kmf}P(C_1=c_1|M=m,F=f).P(C_2=c_2|M=m,F=f)\\
			&.P(D_1=1|M=m,F=f,C_1=c_1).P(D_2=0|M=m,F=f,C_2=c_2)
		\end{aligned}
	\end{equation}

	and
	\begin{equation}
		\begin{aligned}[b]
			\big(C_3^{(i)} & = c_3, C_4^{(i)} = c_4,... ,C_k^{(i)} = c_k, D_3^{(i)}, D_4^{(i)},...,D_k^{(i)}|M^{(i)}=m,F^{(i)}=f\big)\\
			& = P(C_3 = c_3|M^{(i)}=m,F^{(i)}=f)....P(C_k = c_k|M^{(i)}=m,F^{(i)}=f)\\
			&.P(D_3|M^{(i)}=m,F^{(i)}=f,C_3 = c_3)...P(D_k|M^{(i)}=m,F^{(i)}=f,C_k = c_k)
		\end{aligned}
	\end{equation}

	\begin{equation}
		E_{\boldsymbol{\psi}^{(k)}}\{log P_{\boldsymbol{\alpha}}\big(Z\big)|Y\} = \frac{1}{10000} \sum_{j=1}^{10000}\bigg[-log B\big(\boldsymbol{\alpha}\big) + \big(\alpha_{m_i,f_i} - 1\big)\sum_{m=0}^{2}\sum_{f=0}^{2}logZ^{(j)}_{kmf}\bigg]
	\end{equation}
	Then we can define $Q_{MC}(\boldsymbol{\psi};\boldsymbol{\psi}^{(k)})$ as follows
	\begin{equation}
		Q_{MC}(\boldsymbol{\psi},\boldsymbol{\psi}^{(k)}) = \hat{E}_{\boldsymbol{\psi}^{(k)}}\{log P_{\boldsymbol{\theta}}\big(Y|Z\big)|Y\} + E_{\boldsymbol{\psi}^{(k)}}\{log P_{\boldsymbol{\alpha}}\big(Z\big)|Y\}
	\end{equation}
	
	\subsubsection*{\textbf{M Step:}}
	For the maximization step, we maximize $Q_{MC}\big(\boldsymbol{\psi};\boldsymbol{\psi}^{(k)}\big)$ which is the sum of above two equations with respect to the parameters \\
	$\boldsymbol{\psi}$ = \big($\delta, R_1, R_2, R_{im}, S_1, S_1, \alpha_{00}, \alpha_{01}, \alpha_{02}, \alpha_{10}, \alpha_{11}, \alpha_{12}, \alpha_{20}, \alpha_{21}, \alpha_{22} $\big)\\
	Note that the first term of $Q_{MC}\big(\boldsymbol{\psi};\boldsymbol{\psi}^{(k)}\big)$ is only related to $\boldsymbol{\theta}$, and the second term of $Q_{MC}\big(\boldsymbol{\psi};\boldsymbol{\psi}^{(k)}\big)$ is only related to $\boldsymbol{\alpha}$. Therefore, to maximize $Q_{MC}\big(\boldsymbol{\psi};\boldsymbol{\psi}^{(k)}\big)$, we can separately maximize the first term with respect to $\boldsymbol{\theta}$, and maximize the second term with respect to $\boldsymbol{\alpha}$.
	
	\subsection{Applying Importance sampling based on $MCEM_{DSP}$ algorithm}
	It is very time consuming to generate $Z_{mf}$ from posterior distribution using $MCEM_{DSP}$, so we propose to use importance sampling method as the solution for it. \\
	Let $k$ be the iteration number of EM algorithm. \\
	Then up to $10^{th}$ iteration we generate $Z_{mf}$ using the posterior distribution $f(Z|Y;\boldsymbol{\psi}^{(k)})$ as in the $MCEM_{DSP}$ method. At $k \geq 11$ iterations, we'll use the importance weights to calculate $Q_{MC}(\psi;\psi^k)$ as follows.\\
	\begin{equation}
		w_t^{(k)} = \frac{L(\hat{\boldsymbol{\psi}}^{(k)} | Z_{10}^{(t)},Y)}{L(\hat{\boldsymbol{\psi}}^{(0)} | Z_{10}^{(t)},Y)} = \frac{P(Y|Z_{10}^{(t)};\boldsymbol{\theta}^k).P(Z_{10}^{(t)};\boldsymbol{\alpha}^k) }{P(Y|Z_{10}^{(t)};\boldsymbol{\theta}^{10}).P(Z_{10}^{(t)};\boldsymbol{\alpha}^{10}) } = \frac{P(Y|Z_{10}^{(t)};\boldsymbol{\theta}^k).\pi^{\boldsymbol{\alpha}^k}(Z_{10})}{P(Y|Z_{10}^{(t)};\boldsymbol{\theta}^{10}).\pi^{\boldsymbol{\alpha}^{10}}(Z_{10})}
	\end{equation}
	Here, $P(Y|Z_{10}^{(t)};\boldsymbol{\theta}^k)$ the probability density function of Y conditional on $t^{th}$ monte carlo sample of Z at $10^{th}$ iteration with parameter $\boldsymbol{\theta}=\boldsymbol{\theta}^k$ the values from the last importance sampling based $MCEM_{DSP}$ iteration for the parameters $\delta, R_1, R_2, R_{im}, S_1, S_2$.\\
	Then we can compute the Q function in E step as\\
	\begin{equation}
		Q_{MCIm}(\boldsymbol{\psi};\boldsymbol{\psi}^k) = \frac{\sum_{t=1}^{m}w_t^{(k)} \ln f(Z_t,Y | \boldsymbol{\psi})}{\sum_{t=1}^{m}w_t^{(k)}} 
	\end{equation}
	
	For maximization, since $\sum_{t=1}^{k}w_t^{k}$ does not depend on unknown parameters, we can simply maximize the numerator with respect to $\boldsymbol{\theta}$ and $\boldsymbol{\alpha}$ to update $\psi$.
	\begin{equation}
		\sum_{t=1}^{m}w_t^{(k)} \ln f(Z_t,Y | \psi) = \sum_{t=1}^{m}w_t^{(k)} \ln f(Y|Z_t;\boldsymbol{\theta}) + \sum_{t=1}^{m}w_t^{(k)} \ln f(Z_t| \boldsymbol{\alpha}) 
	\end{equation} 
	
	The first term of the numerator of $Q_{MCIm}$ is only related to $\boldsymbol{\theta}$ and the second term of $Q_{MCIm}$ is only related to $\boldsymbol{\alpha}$. So we can maximize $Q_{MCIm}$ by separately maximizing the first part with respect to $\boldsymbol{\theta}$ and maximizing the second part with respect to $\boldsymbol{\alpha}$. 
	\subsection{Hypothesis Testing }
	
	To test whether the candidate marker has association effect, imprinting effect, and maternal effect with the disease. We need to consider the follower four models.\\
	\begin{enumerate}
		\item Full model\\
		In this model, we consider child's own genotype effect, imprinting effect, and maternal effect are all possibly related to the disease. \\
		$$P(D=1|M=m,F=f,C=c) = \delta R_{1}^{I(c=1)} R_{2}^{I(c=2)} R_{im}^{I(c=1_m)} S_{1}^{I(m=1)} S_{2}^{I(m=2)}$$ \\
		
		\item Null model\\
		For null model, we consider there is no effect from child's own genotype effect, imprinting effect, or maternal effect in the model, i.e., the disease penetrance is the same as the phenocopy rate.\\
		$$P(D=1|M=m,F=f,C=c) = \delta$$ \\
		\item Non-Imprinting effect model\\
		In this model there is no imprinting effect, i.e., $R_{im} = 1$. \\
		$$P(D=1|M=m,F=f,C=c) = \delta R_{1}^{I(c=1)} R_{2}^{I(c=2)} S_{1}^{I(m=1)} S_{2}^{I(m=2)}$$ \\
		\item Non-Maternal effect model\\
		In this model, there is no maternal effect, i.e., $S_1$ and $S_2$ are both equal to 1. \\
		$$P(D=1|M=m,F=f,C=c) = \delta R_{1}^{I(c=1)} R_{2}^{I(c=2)} R_{im}^{I(c=1_m)}$$ \\
	\end{enumerate}
	
	\subsubsection*{Association effect}
	For testing association effect, we check the following hypothesis. \\
	\begin{center}
		$H_0 : R_1 = R_2 = R_{im} = S_1 = S_2 = 1$ (NULL model)\\
		vs.\\
		$H_a$ : at least one of these parameters is not equal to 1 (Full model )\\	
	\end{center}
	The test statistic is,\\
	\begin{equation}
		T_1 = \frac{1}{10000}\sum_{t=1}^{10000}2\big[log P_{FULL}\big(Y|Z^{(t)}\big) - logP_{NULL}\big(Y|Z^{(t)}\big)\big]  	
	\end{equation}
	
	Here, $P_{FULL}\big(Y|Z^{(t)}\big)$ is the likelihood value for the full model and $P_{NULL}\big(Y|Z^{(t)}\big)$ is the likelihood for NULL model.\\
	When there are additional siblings, the test statistic asymptotically follows $\chi^2_5$, while when there are no additional siblings, the test statistic asymptotically follow $\chi^2_6$. This is because in $\log P(Y|Z)$ formula, the discordant sibpair part is free of all the parameters in $\boldsymbol{\theta}$ ($\delta(1-\delta)$ appear in both numerator and denominator and can be cancelled), while additional sibling part includes $\delta$.
	\begin{equation}
		\begin{aligned}[b]
			& log P(Y|Z)  = \sum_{i=1}^{n} \Biggl\{ log \Bigg[ \frac{  P(M^{(i)}=m,F^{(i)}=f,C_1^{(i)}=c_1,C_2^{(i)}=c_2,D_1^{(i)}=1,D_2^{(i)}=0)}{P(D^{(i)}_1=1,D^{(i)}_2=0)}\Bigg] \\ 
			& + \log \bigg(P\big(C_3^{(i)} = c_3, C_4^{(i)} = c_4,...,C_k^{(i)} = c_k, D_3^{(i)}, D_4^{(i)},...,D_k^{(i)}|M^{(i)}=m,F^{(i)}=f\big)\bigg)\Biggr\}\\
			& = \sum_{i=1}^{n} \Biggl\{ log \Bigg[ \frac{ \mu_{mf}.P(C_1 = c_1|M=m,F=f).P(C_2 = c_1|M=m,F=f)\delta(1-\delta) }{\sum_{m,f,c_1,c_2} \mu_{mf}.P(C_1 = c_1|M=m,F=f).P(C_2 = c_1|M=m,F=f)\delta(1-\delta)}\Bigg] \\ 
			& + \log \bigg(P(C_3 = c_3|M=m,F=f)....P(C_k = c_k|M=m,F=f)\\
			&.P(D_3|M=m,F=f,C_3 = c_3)...P(D_k|M=m,F=f,C_k = c_k)\bigg)\Biggr\}\\
			& = \sum_{i=1}^{n} \Biggl\{ log \Bigg[ \frac{ \mu_{mf}.P(C_1 = c_1|M=m,F=f).P(C_2 = c_1|M=m,F=f)}{\sum_{m,f,c_1,c_2} \mu_{mf}.P(C_1 = c_1|M=m,F=f).P(C_2 = c_1|M=m,F=f)}\Bigg] \\
			& + \log \bigg(P(C_3 = c_3|M=m,F=f)....P(C_k = c_k|M=m,F=f)\\
			&.\delta^{\sum_{l = 3}^{k}D_l}(1 - \delta)^{k-2-\sum_{l=3}^{k}D_l}\bigg)\Biggr\}\\ 
		\end{aligned}
	\end{equation}
	
	\subsubsection*{Imprinting effect}
	To test the imprinting effect, we consider the following hypothesis.\\
	\begin{center}
		$H_0 : R_{im} = 1$ (Non-imprinting model)\\
		vs. \\
		$H_a : R_{im} \neq 1$ (Full model)\\
	\end{center}
	The test statistic is\\
	\begin{equation}
		T_2 = \frac{1}{10000}\sum_{t=1}^{10000}2\big[log P_{FULL}\big(Y|Z^{(t)}\big) - logP_{Non-Imprinting}\big(Y|Z^{(t)}\big)\big] \sim \chi^2_1   
	\end{equation}
	under null hypothesis.
	\subsubsection*{Maternal effect}
	There exists maternal effect means $S_1 \neq 1$ or $S_2 \neq 1$. Thus, the hypothesis for the maternal effect is\\
	\begin{center}
		$H_0 : S_1 = S_2 = 1$ (Non-Maternal effect model)\\
		vs.\\
		$H_a$ : $S_1$ or $S_2$ is not equal to 1 (Full model)\\
	\end{center}	
	The test statistic for check the above hypothesis is,\\
	\begin{equation}
		T_3 = \frac{1}{10000}\sum_{t=1}^{10000}2\big[log P_{FULL}\big(Y|Z^{(t)}\big) - logP_{Non-Maternal}\big(Y|Z^{(t)}\big)\big] \sim \chi^2_2 
	\end{equation}
	under null hypothesis.\\
	\vspace{0.5cm}For all of above three cases, p-value is calculated based on the test statistic and then compared with significance level.\\ 

\section{Simulation}	
	To examine the performance of $MCEM_{DSP}$ method, we consider eight disease models and eight scenarios.\\
	\begin{table}[!htb]
		\caption{Eight simulation settings of relative risks and eight scenarios with three factors}
		\centering
		\begin{tabular}{c c c c c c c c c c c c c}
			\hline \hline
			\cline{1-6} \cline{8-11}  
			\multicolumn{6}{c}{Disease Model} & & \multicolumn{4}{c}{Factor values}\\
			\cline{1-6} \cline{8-11}
			& \multicolumn{5}{c}{Parameters} &  & & \multicolumn{3}{c}{Scenarios}\\
			\cline{2-6} \cline{8-11}
			Settings & $R_1$ & $R_2$ & $R_{im}$ & $S_1$ & $S_2$ & &Settings & MAF & PREV & HWE\\ 
			$1$ & $1$ & $1$ & $1$ & $1$ & $1$ & & $1$ & $0.1$ & $0.05$ & $0$\\
			$2$	& $2$ &	$3$	& $1$ & $1$ & $1$ & & $2$	& $0.3$ & $0.05$ & $0$\\
			$3$	& $1$ &	$3$ & $1$ & $1$	& $1$ & & $3$	& $0.1$ & $0.15$ & $0$\\
			$4$	& $1$ & $3$	& $1$ & $2$	& $2$ & & $4$	& $0.3$ & $0.15$ & $0$\\
			$5$	& $1$ & $3$	& $3$ & $1$	& $1$ & & $5$	& $0.1$ & $0.05$ & $1$\\
			$6$	& $3$ & $3$	& $1/3$ & $1$ & $1$ & & $6$	& $0.3$ & $0.05$ & $1$\\
			$7$	& $1$ & $3$	& $3$ & $2$ & $2$ & & $7$	& $0.1$ & $0.15$ & $1$\\
			$8$	& $3$ & $3$	& $1/3$ & $2$ & $2$ & & $8$	& $0.3$ & $0.15$ & $1$\\
		\end{tabular}
	\end{table}
	In table 3.1, the first setting for the disease model corresponds to null model with no genetic effect. The settings 2 and 3 correspond to the disease models that there is neither imprinting nor maternal effect. Setting 4 only has maternal effect. Settings 5 and 6 only have imprinting effect. Settings 7 and 8 have both imprinting and maternal effects. Under each model, we investigate eight different scenarios in which there are three factor values: minor allele frequency (MAF) $\{0.1,0.3\}$, disease prevalence $P(D = 1)$ (PREV) $\{0.05 (rare), 0.15 (common)\}$ and condition for Hardy-Weinberg equilibrium (HWE) $\{1 = Yes, 0 = No\}$. Note that, $p$ is the frequency for the minor allele and $(1 - p)$ is wild allele frequency. As a result of population HWE, allelic exchangeability and mating symmetry will be implied. In cases that HWE is violated, we use a common assumption in population genetics and assume that a common cause of non-random matings is inbreeding. Then, in such cases, the probability of genotypes containing 0, 1, and 2 minor alleles are $(1 - p)^2(1 - \zeta) + (1 - p)\zeta$, $2p(1 - p)(1 - \zeta)$ and $p^2(1 - \zeta) + p\zeta$, respectively, where $\zeta$ is the inbreeding parameter. When HWE holds, $\zeta = 0$. $\zeta$ is set to be 0.1 and 0.3 for males and females, respectively, when HWE does not hold. Note that with the specification of each scenario and a disease model, the penetrance probability is fully specified. \\
	We generate 500 replications under each of these 64 settings. To check the influence of sample size, each replicate consists of 100 or 500 discordant sibpair families. Firstly, parental genotypes are generated based on MAF and HWE. Then, the genotypes of their proband children are created according to the transmission probability assuming no recombination. Affection status D of the probands are determined by a Bernoulli trial, with the success probability calculated based on disease penetrance model. A family with an affected child and an unaffected child is recruited as a discordant sibpair family.  The process of generating M, F,C, and D is repeated until we have collected sufficient numbers of discordant sibpair families to meet the preset sample size. We further simulate the genotype and disease status of a third child for each family as an additional sibling. We denote the data without additional sibling as \enquote{DS}, and denote the data with one additional sibling in each family as \enquote{DS+1}.

	\section{Results}
	
	\subsection{Type I error and power}
	Under eight disease models and eight scenarios, we compare the type I error and power of $MCEM_{DSP}$ and $LIME_{DSP}$ for data types \enquote{SD} and \enquote{SD+1}.  Each figure from Figures 3.1 and 3.2 represent first two scenarios for sample size 100 (Appendix figures 1 - 6 are for the other scenarios). In the figures, the three rows represent association effect, imprinting effect, and maternal effect. Within each row, there are eight clusters of four bars representing the 8 disease models. The four bars in each cluster represent type I error (denoted as \enquote{E}) or power (denoted as \enquote{P}) of $LIME_{DSP}$ and $MCEM_{DSP}$ methods on \enquote{SD} data, and $LIME_{DSP}$ and $MCEM_{DSP}$ methods on \enquote{SD+1} data, respectively. According to the disease models, type I error for association effect can be obtained from disease model 1; type I error for imprinting effect can be obtained from disease model 1-4; type I error for maternal effect can be obtained from disease models 1,2, 3, 4, and 6. Powers can be obtained in other disease models. \\ 
	\begin{figure}[ht!]
		\centering
		\includegraphics[width=\linewidth]{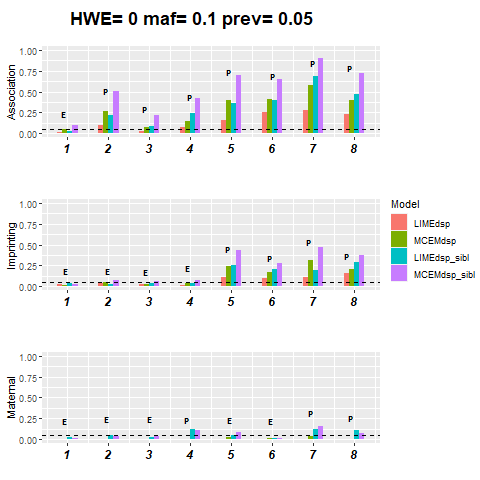}
		\caption{Bar charts for compare type I error and power of $MCEM_{DSP}$ and $LIME_{DSP}$ methods with and without additional siblings when HWE = 0, maf = 0.1 and prev = 0.05 with 100 families}
	\end{figure}
	\begin{figure}[ht!]
		\includegraphics[width=\linewidth]{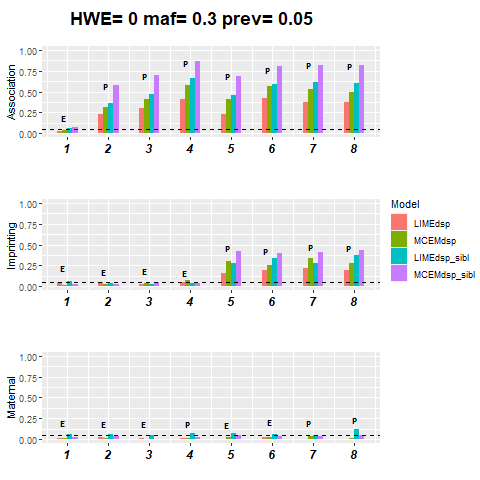}
		\caption{Bar charts for compare type I error and power of $MCEM_{DSP}$ and $LIME_{DSP}$ methods with and without additional siblings when HWE = 0, maf = 0.3 and prev = 0.05 with 100 families}
	\end{figure}
	
	The figures show that $MCEM_{DSP}$ method can control all the type I error at around 0.05 nominal values, even when HWE does not hold. This means $MCEM_{DSP}$ is very robust to violation of HWE assumption. We can also see that no matter whether the discordant sibpair families have additional siblings or not, $MCEM_{DSP}$ method is generally more powerful than $LIME_{DSP}$. As expected, the powers of the methods with additional siblings are higher than the methods without additional siblings. We can also see that for some settings and scenarios, such as setting 5 and 7 when disease prevalence is 0.05, the imprinting effect power of $MCEM_{DSP}$ for the data without additional siblings can be even higher than that of $LIME_{DSP}$ for the data with one additional sibling. The figures show that the power to detect maternal effects are generally very low for both $MCEM_{DSP}$ and $LIME_{DSP}$ methods, especially for the data without additional siblings. The main reason is in discordant sibpair design the affected proband and unaffected proband share the same mother leading to little contrast to detect maternal effect.\\
	
	\begin{figure}[ht!]
		\centering
		\includegraphics[width=\linewidth]{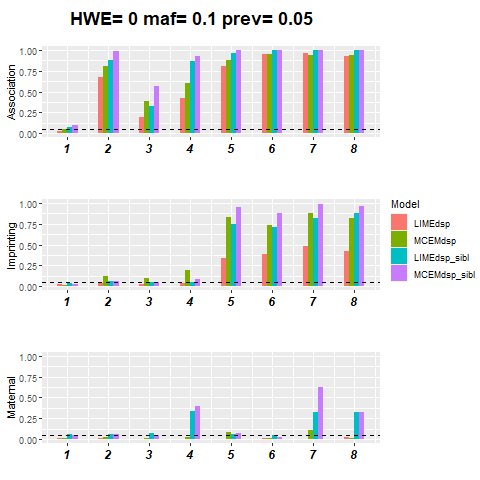}
		\caption{Bar charts for compare type I error and power of $MCEM_{DSP}$ and $LIME_{DSP}$ methods with and without additional siblings when HWE = 0, maf = 0.1 and prev = 0.05 with 500 families}
	\end{figure}
	\begin{figure}[ht!]
		\includegraphics[width=\linewidth]{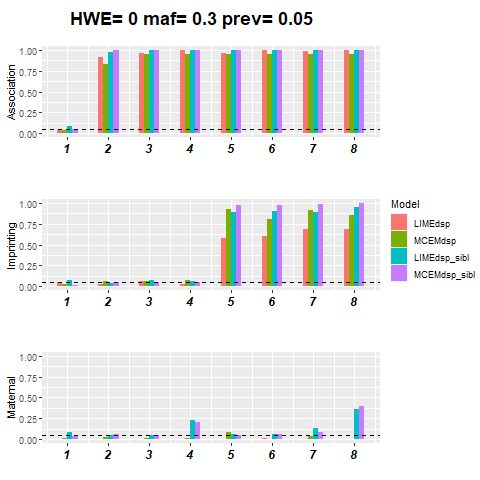}
		\caption{Bar charts for compare type I error and power of $MCEM_{DSP}$ and $LIME_{DSP}$ methods with and without additional siblings when HWE = 0, maf = 0.3 and prev = 0.05 with 500 families}
	\end{figure}
	
	Each figure from Figures 3.3 and 3.4 represent first two scenarios when sample size is 500 (Appendix figures 7 - 12 are for the other scenarios). We can see that the powers are generally higher compared to that when sample size is 100, but the results of power comparison between different methods and data types are similar. The type I error, however, show different patterns. Type I errors are generally close to nominal value 0.05, except for imprinting effect when minor allele frequency is low and there are no additional siblings, which implies that under this situation, $MCEM_{DSP}$ algorithm cannot converge to the true parameter values, as small minor allele frequency might lead to flat complete log likelihood function. This inflation of type I error might be also related to the confounding between imprinting effect and maternal effect. As we mentioned, maternal effect can mimic paternal imprinting effect. $MCEM_{DSP}$ cannot differentiate the two confounding epigenetic effects in disease model 4 $((R_1, R_2, R_{im}, S_1, S_2)=(1,3,1,2,2))$, where there is maternal effect, but no imprinting effect under the scenario MAF=0.1, sample size = 500, and data type=\enquote{SD}. 
	\clearpage
	\subsection{Relative bias of the parameter estimation}
	We define the relative bias as $\frac{(\hat{\theta} - \theta)}{\theta}$. The relative bias is calculated under each scenario and disease model for discordant sib-pair families with and without additional siblings, and shown in boxplots 3.5 and 3.6 (Appendix figures 13 - 18 are for the other scenarios) when sample size is 100. There are five rows in each of these boxplot figure, representing the relative bias for $R_1$, $R_2$, $R_{im}$, $S_1$, and $S_2$, respectively. The eight clusters of boxplots represent the eight disease models. The four boxplots in each cluster represent the results for LIMEdsp and MCEMdsp for \enquote{SD} data and LIMEdsp and MCEMdsp for \enquote{SD+1} data, respectively.\\
	We can see that the relative bias medians for the parameters are all around 0 for both $MCEM_{DSP}$ and $LIME_{DSP}$ under all settings. The interquartile range for $R_1$, $R_2$, $R_{im}$, $S_2$ are generally
	smaller for $MCEM_{DSP}$. The interquartile range for $S_1$ is bigger for $MCEM_{DSP}$ when there are no additional sibling, but it becomes smaller than $LIME_{DSP}$ when there are additional siblings.
	
	\begin{figure}[ht!]
		\centering
		\includegraphics[width=0.9\textwidth]{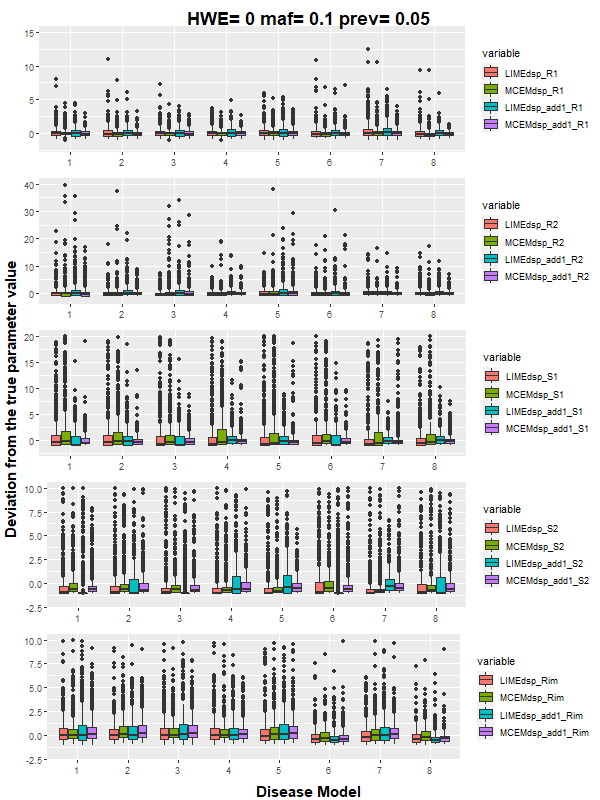}
		\caption{Box plots for biases of parameters from $MCEM_{DSP}$ and $LIME_{DSP}$ methods when HWE = 0, maf = 0.1 and PREV = 0.05 with 100 families}
	\end{figure}
	\begin{figure}[ht!]
		\centering
		\includegraphics[width=0.9\textwidth]{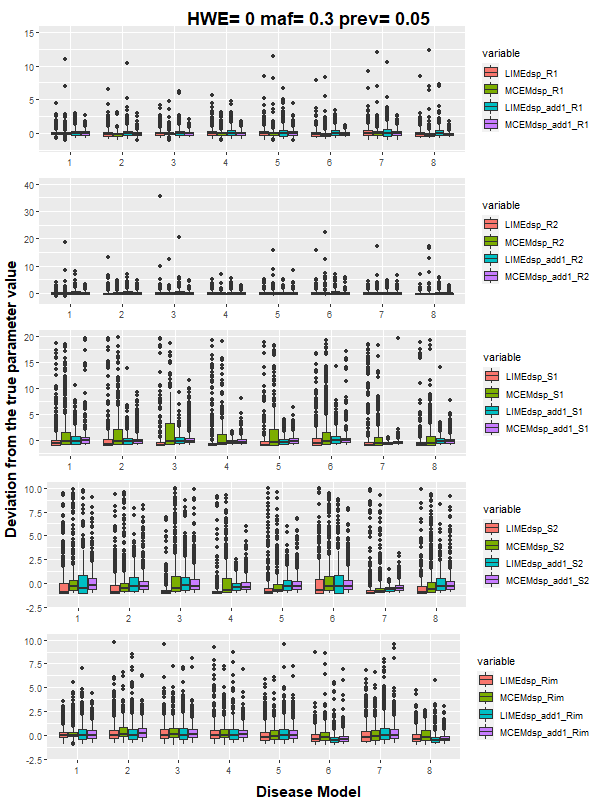}
		\caption{Box plots for biases of parameters from $MCEM_{DSP}$ and $LIME_{DSP}$ methods when HWE = 0, maf = 0.3 and PREV = 0.05 with 100 families}
	\end{figure}
	\begin{figure}[ht!]
		\centering
		\includegraphics[width=0.9\textwidth]{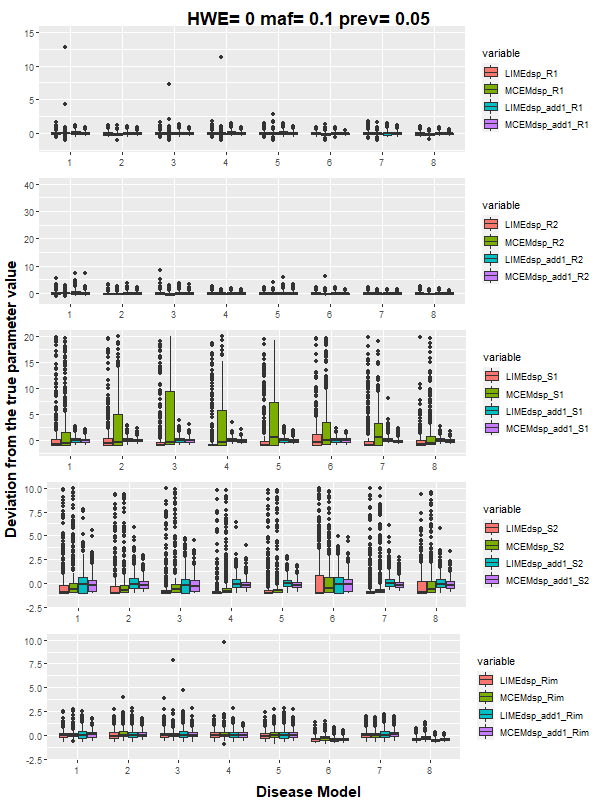}
		\caption{Box plots for biases of parameters from $MCEM_{DSP}$ and $LIME_{DSP}$ methods when HWE = 0, maf = 0.1 and PREV = 0.05 with 500 families}
	\end{figure}
	\begin{figure}[ht!]
		\centering
		\includegraphics[width=0.9\textwidth]{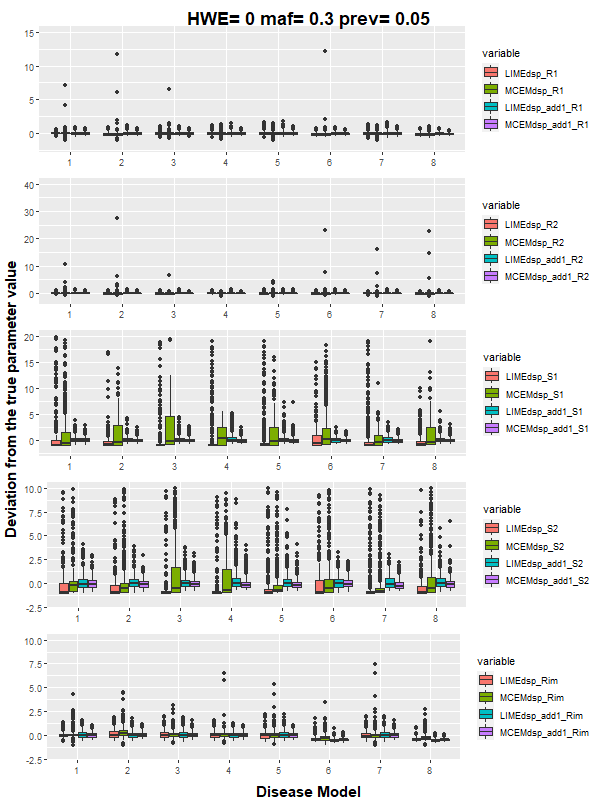}
		\caption{Box plots for biases of parameters from $MCEM_{DSP}$ and $LIME_{DSP}$ methods when HWE = 0, maf = 0.3 and PREV = 0.05 with 500 families}
	\end{figure}
	The Figures 3.7 and 3.8 represent first two scenarios when sample size is 500 (Appendix figures 19 - 24 are for the other scenarios). The relative bias comparison results between the two methods are similar for the two samples sizes, but the interquartile ranges are all smaller when sample size is larger. We can also see that both $MCEM_{DSP}$ and $LIME_{DSP}$ have smaller  proportion of wild estimation for the data with larger sizes or with additional siblings.
	\clearpage
	\subsection{Importance Sampling method}
	Though $MCEM_{DSP}$ estimation gives a tractable solution to the problem arising when E-step is not available in closed form, the computation time cost is a major issue when implementing the $MCEM_{DSP}$ routine. \\
	Let's consider the first disease model in Table 3.1 as an example.\\
	\begin{table}[!htb]
		\caption{Computational time comparison for $LIME_{DSP}$, $MCEM_{DSP}$ and importance sampling based $MCEM_{DSP}$ methods }
		\centering
		\begin{tabular}{|l|l|l|l|l|} 
			\hline
			\small{Method} & \small{Resource} & \small{Nodes} & \small{Replicates} & \small{Computational time}\\
			\hline
			$\small{LIME_{dsp}}$ & \small{HPCC} & \small{nodes = 1} & \small{500} & \small{10 minutes}\\
			\cline{1-1} \cline{4-5}
			$\small{MCEM_{dsp}}$ & \small{CentOS 7.4} & \small{cores = 500} & \small{500} & \small{10 hours}\\
			\cline{1-1} \cline{4-5}
			\small{Importance Sampling} & & & \small{500} & \small{4 hour}\\
			\hline
		\end{tabular}
	\end{table}
	We apply importance sampling based $MCEM_{DSP}$ method for eight disease models when Hardy-Weinberg equilibrium (HWE) is 0, disease prevalence (prev) is 0.05 and minor allele frequency (MAF) is set to be 0.1 with sample size 500 for the families with one additional sibling. The simulation shows that, the type I error and the power of importance sampling based $MCEM_{DSP}$ are comparable to $MCEM_{DSP}$ method. For some disease models, the power of the importance sampling based $MCEM_{DSP}$ is a little lower than $MCEM_{DSP}$, but it is still
	generally higher than $LIME_{DSP}$ method. What's more, the important sampling based $MCEM_{DSP}$ is much more time efficient than $MCEM_{DSP}$ method.
	\begin{figure}
		\centering
		\includegraphics[width=1.0\textwidth]{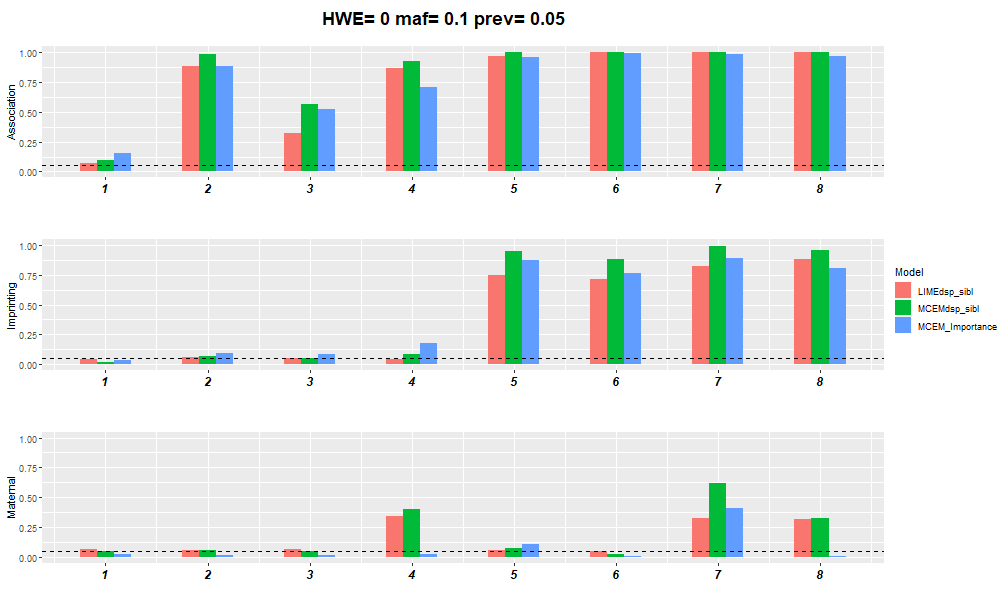}
		\caption{Comparison of power and type I error of $LIME_{DSP}$, $MCEM_{DSP}$ and importance sampling methods }
	\end{figure} 

\section{Real Data Analysis}
To illustrate application of $MCEM_{DSP}$ method we apply it to Framingham Heart Study (FHS). FHS is a long-term, ongoing cardiovascular risk study to examine the epidemiology of cardiovascular diseases on cohorts of residents of Framingham, Massachusetts. This study is considered as the first prospective study on cardiovascular disease which identified risk factors and their joint effects. In FHS several cardiovascular disease conditions were considered including coronary heart disease, stroke, hypertension, peripheral arterial disease, and congestive heart failure. In this study, we focus on hypertension, a multifactorial complex trait, which can increase the risk of coronary heart disease. A person is categorized as hypertensive, if her/his systolic blood pressure is $\geq$ 140mmHg or diastolic blood pressure is $\geq$ 90mmHg or he/she has taken medication to control blood pressure. In this study, we considered 48071 SNPs in 22 autosomal chromosomes for 263 discordant sib-pair families with 229 additional siblings. As in our simulations, $MCEM_{DSP}$ can control type I error well for all the settings when there are additional siblings, we believe $MCEM_{DSP}$ is also valid in this real data analysis. \\
	Many of the top SNPs identified to be associated with the hypertensive trait by $MCEM_{DSP}$ (Table 3.4) have been previously showed in the literature related to hypertension, cardiovascular disorder or other complex disease. Specifically, SNP rs12117125, inhabiting in the gene FCGR2A on chromosome 1,is found to be associated with Kawasaki disease. It has been identified in association with damage to the coronary arteries, making it a notable contributor to pediatric acquired heart disease. The potential impact on the coronary arteries underscores the significance of addressing and managing this condition in children.\cite{65}. SNP rs12416299 is another SNP  identified to be associated with blood pressure, presence of hypertension, and age at onset of hypertension \cite{53} . It is located in the gene SORBS1 on chromosome 10. The gene SVEP1 in chromosome 9 (containing rs1327533) is associated with hypertension and type 2 diabetes \cite{54}.\\
	It is worth to discussing several genes found to have an imprinting effect on hypertension (Table 3.4). Previous research suggests that, the gene CD44 in chromosome 11 (containing rs353637) is related to pulmonary arterial hypertension \cite{55}. SNP rs16832560 in gene SOX2-OT in chromosome 3, is also diagnosed as related to pulmonary arterial hypertension which regulates the proliferation, migration, anti-apoptosis, and inflammation of pulmonary artery smooth muscle cells \cite{56}. The gene TRPM8 in chromosome 2 (containing rs7593557) is found to be associated with renin - antiotensin - aldosterone system - mediated hypertension, which subsequently induces small molecule and fluid extravasation, increases plasma Ig levels, and elicits immunosuppression \cite{57}. Also, according to literature, the gene ATP11A in chromosome 13 (containing rs381583) may act as a susceptibility gene for pulmonary fibrosis and an indicator of pulmonary hypertension \cite{58}. \\ 
	When considering the maternal effect, there are several genes that harbor SNPs which are found to have a significant effect on the hypertensive trait by $MCEM_{DSP}$ (Table 3.4). The previous literature shows that, SNP rs4961, inhabiting in gene ADD1 in chromosome 4, is associated with hypertension. It was found to be significantly associated with increased prevalence of peripheral arterial disease and incidence of coronary heart disease among hypertensive individuals \cite{59}. As another example, rs3811515 within gene SPHKAP (chromosome 2) is associated with hypertension \cite{60}. The gene ROBO4 in chromosome 11 (containing rs7129737) is identified that it has an association with pulmonary artery hypertension \cite{61}. Gene DOT1L in chromosome 19 (containing rs12986413) was found to have a strong association with greater systolic and diastolic blood pressure \cite{62}.  However, when considering the top genes, almost all of them have higher minor allele frequency ($\geq 0.2$). A full list of the top SNPs identified by $MCEM_{DSP}$ can be found in Appendix Table 1,2, and 3. \\
	We also apply $LIME_{DSP}$ method to the FHS data, and calculate the Bonferroni method adjusted p-values. We find that at 0.05 significance level, no SNPs have association, imprinting, or maternal effects based on $LIME_{DSP}$ method. $MCEM_{DSP}$, on the other hand, can identify 45 SNPs with association, 4 SNPs with imprinting effect, and 4 SNPs with maternal effect at 0.05 significance level. This result is consistent with our conclusion in simulations that $MCEM_{DSP}$ is generally more powerful than $LIME_{DSP}$. 
	\begin{table}[!htb]
		\caption{Some significant findings of association, imprinting and maternal effects for the Framingham Heart Study data using $MCEM_{DSP}$}
		\centering
		\begin{tabular}{| l| l| l| l| l | l|} 
			\hline
			Effect & SNP & Chr & Position(BP) & Gene & $-log_{10}$(P-value)  \\
			\hline
			Association & rs12117125 & 1 & 161522658 & FCGR2A & 12.3787 \\
			& rs12416299 & 10 & 95542074 & SORBS1 & 15.4775 \\
			& rs1327533 & 9 & 110368883 & SVEP1 & 12.0632 \\
			\hline
			Imprinting & rs353637 & 11 & 35163005 & CD44 & 3.6998 \\
			& rs16832560 & 3 & 181564785 & SOX2-OT & 3.6316 \\
			& rs7593557 & 2 & 233955144 & TRPM8 & 3.5197 \\
			\hline
			Maternal & rs4961 & 4 & 2904980 & ADD1 & 4.4915\\
			& rs3811515 & 2 & 228018313 & SPHKAP & 4.2972 \\
			& rs7129737 & 11 & 124888741 & ROBO4 & 4.1144 \\
			& rs12986413 & 19 & 2170955 & DOT1L & 4.0802 \\
			\hline	
		\end{tabular}
	\end{table}

\section{Concluding Remarks and Summary}

	Both imprinting and maternal effects are important sources of missing heritability in complex human diseases that cannot be explained by genome-wide association studies. Among all the existing methods to detect these two confounded epigenetic effects, almost all the full likelihood -based methods rely on strong yet unrealistic assumptions concerning mating type probabilities to avoid over parameterization. Two partial likelihood – based methods, LIME and $LIME_{DSP}$, can exceptionally overcome the over-parameterization problem without making those assumptions. LIME requires case families and control families, while $LIME_{DSP}$ only requires families with discordant sibpair that is easier to recruit. Empirical and theoretical results have shown their validity and robustness, but it is likely that these partial likelihood – based methods are less efficient than full-likelihood based methods in term of parameter estimation. An MCEM algorithm was developed to find full likelihood maximizer based on case families and control families and was found to be more powerful than LIME. In this study, we further propose an $MCEM_{DSP}$ algorithm to find full likelihood maximizer based on discordant sibpair design. This $MCEM_{DSP}$ algorithm can detect the two epigenetic effects jointly and can accommodate discordant sibpair families with an arbitrary number of additional siblings.\\
	Expectation maximization method is often used for finding full likelihood maximizer when there are missing data or unobservable latent random variables. In our study, we use mating type probabilities (the nuisance parameters) as latent random variables in the EM algorithm to get full likelihood maximizer without making assumptions about them. As the expectation in E-step is not in closed format, monte carlo simulations are used to estimate the expectation. Log likelihood ratio test is used to test association, imprinting effect and maternal effect related to the disease of interest. Extensive simulations under different disease models and scenarios demonstrate that the $MCEM_{DSP}$ method can generally control type I error under most scenarios and is more powerful than $LIME_{DSP}$. For some simulation settings, the power gain by using $MCEM_{DSP}$ method instead of $LIME_{DSP}$ method can be even higher than the power gain by recruiting one more additional sibling in each family based on $LIME_{DSP}$. In addition, the parameter estimation based on $MCEM_{DSP}$ method has similar relative bias and generally smaller standard error compared to $LIME_{DSP}$ method. To illustrate the utility of $MCEM_{DSP}$ method, we apply it to Framingham Heart Study data. The results show that many of our findings are consistent with those in the literature, but potential novel genes also emerged.\\
	Despite its advantages, $MCEM_{DSP}$ has several limitations. Firstly, simulations show that the empirical type I error for imprinting effect is inflated when the data include 500 discordant sibpairs without any additional siblings and minor allele frequency is low (0.01). This is probably because the low minor allele frequency leads to a flatter expectation function in the E-step of the algorithm, such that the M-step cannot identify the global likelihood maximizer. Secondly, $MCEM_{DSP}$ algorithm takes much longer time than $LIME_{DSP}$, due to the time cost in generating monte carlo samples in each iteration. To alleviate this problem, we propose an importance sampling based $MCEM_{DSP}$ algorithm, which can reweight and repeatedly use the monte carlo samples in E-step to greatly reduce the time consumption. Besides, simulations show that importance sampling based $MCEM_{DSP}$ algorithm has similar performance as the original $MCEM_{DSP}$.\\
	In future works, we want to explore whether any modification of current $MCEM_{DSP}$ algorithm can correct the type I error inflation under some situations as in the simulations. In addition, we also want to check the possibility of applying $MCEM_{DSP}$ to the data of discordant sibpair families with father’s genotype missing. It is well known that fathers are usually much harder to recruit than mothers for a genetic study, and thus a study design with discordant sibpair and their mother may be easier to meeting its target sample size. Nevertheless, with father’s genotype missing, when both mother’s and child’s are heterogeneous, we cannot determine the parental origin of child’s minor allele, thus the related penetrance probability calculation will become more complex. $LIME_{DSP}$ actually fails to deal with this type of data, as the related partial likelihood will become no longer free of the nuisance parameters to get rid of the over-parameterization problem. As $MCEM_{DSP}$ algorithm is based on full likelihood, we expect that it can accommodate this type of incomplete family data.\\

\newpage
\bibliographystyle{unsrtnat}
\bibliography{Reference_new}

\begin{thebibliography}{51}
\providecommand{\natexlab}[1]{#1}
\providecommand{\url}[1]{\texttt{#1}}
\expandafter\ifx\csname urlstyle\endcsname\relax
  \providecommand{\doi}[1]{doi: #1}\else
  \providecommand{\doi}{doi: \begingroup \urlstyle{rm}\Url}\fi

\bibitem[Manolio et~al.(2009)Manolio, Collins, Cox, Goldstein, Hindorff,
  Hunter, McCarthy, Ramos, Cardon, Chakravarti, et~al.]{1}
Teri~A Manolio, Francis~S Collins, Nancy~J Cox, David~B Goldstein, Lucia~A
  Hindorff, David~J Hunter, Mark~I McCarthy, Erin~M Ramos, Lon~R Cardon,
  Aravinda Chakravarti, et~al.
\newblock Finding the missing heritability of complex diseases.
\newblock \emph{Nature}, 461\penalty0 (7265):\penalty0 747--753, 2009.

\bibitem[Yang et~al.(2017)Yang, Zeng, Goddard, Wray, and Visscher]{2}
Jian Yang, Jian Zeng, Michael~E Goddard, Naomi~R Wray, and Peter~M Visscher.
\newblock Concepts, estimation and interpretation of snp-based heritability.
\newblock \emph{Nature genetics}, 49\penalty0 (9):\penalty0 1304--1310, 2017.

\bibitem[Guo et~al.(2018)Guo, Hirschhorn, and Dauber]{3}
Michael~H Guo, Joel~N Hirschhorn, and Andrew Dauber.
\newblock Insights and implications of genome-wide association studies of
  height.
\newblock \emph{The Journal of Clinical Endocrinology \& Metabolism},
  103\penalty0 (9):\penalty0 3155--3168, 2018.

\bibitem[Peters(2014)]{4}
Jo~Peters.
\newblock The role of genomic imprinting in biology and disease: an expanding
  view.
\newblock \emph{Nature Reviews Genetics}, 15\penalty0 (8):\penalty0 517--530,
  2014.

\bibitem[Dupont et~al.(2009)Dupont, Armant, and Brenner]{5}
Cath{\'e}rine Dupont, D~Randall Armant, and Carol~A Brenner.
\newblock Epigenetics: definition, mechanisms and clinical perspective.
\newblock In \emph{Seminars in reproductive medicine}, volume~27, pages
  351--357. {\copyright} Thieme Medical Publishers, 2009.

\bibitem[Kohda(2013)]{6}
Takashi Kohda.
\newblock Effects of embryonic manipulation and epigenetics.
\newblock \emph{Journal of human genetics}, 58\penalty0 (7):\penalty0 416--420,
  2013.

\bibitem[Wilkinson et~al.(2007)Wilkinson, Davies, and Isles]{9}
Lawrence~S Wilkinson, William Davies, and Anthony~R Isles.
\newblock Genomic imprinting effects on brain development and function.
\newblock \emph{Nature Reviews Neuroscience}, 8\penalty0 (11):\penalty0
  832--843, 2007.

\bibitem[Ferguson-Smith(2011)]{10}
Anne~C Ferguson-Smith.
\newblock Genomic imprinting: the emergence of an epigenetic paradigm.
\newblock \emph{Nature Reviews Genetics}, 12\penalty0 (8):\penalty0 565--575,
  2011.

\bibitem[Jirtle and Weidman(2007)]{12}
Randy~L Jirtle and Jennifer~R Weidman.
\newblock Imprinted and more equal.
\newblock \emph{Am Sci}, 95:\penalty0 143--149, 2007.

\bibitem[Lobo(2008)]{13}
I~Lobo.
\newblock Genomic imprinting and patterns of disease inheritance.
\newblock \emph{Nat. Educ}, 1\penalty0 (5), 2008.

\bibitem[Allis and Jenuwein(2016)]{14}
C~David Allis and Thomas Jenuwein.
\newblock The molecular hallmarks of epigenetic control.
\newblock \emph{Nature Reviews Genetics}, 17\penalty0 (8):\penalty0 487--500,
  2016.

\bibitem[Scharfmann and Shield(2007)]{15}
Raphael Scharfmann and Julian~PH Shield.
\newblock \emph{Development of the pancreas and neonatal diabetes}, volume~12.
\newblock Karger Medical and Scientific Publishers, 2007.

\bibitem[Anthony(1993)]{11}
J~F~Griffiths Anthony.
\newblock \emph{An introduction to genetic analysis}.
\newblock WH Freeman, 1993.

\bibitem[Haig(2004)]{16}
David Haig.
\newblock Evolutionary conflicts in pregnancy and calcium metabolism—a
  review.
\newblock \emph{Placenta}, 25:\penalty0 S10--S15, 2004.

\bibitem[Palmer et~al.(2008)Palmer, Mallery, Turunen, Hsieh, Peltonen,
  Lonnqvist, Woodward, and Sinsheimer]{17}
Christina~GS Palmer, Erin Mallery, Joni~A Turunen, Hsin-Ju Hsieh, Leena
  Peltonen, Jouko Lonnqvist, J~Arthur Woodward, and Janet~S Sinsheimer.
\newblock Effect of rhesus d incompatibility on schizophrenia depends on
  offspring sex.
\newblock \emph{Schizophrenia research}, 104\penalty0 (1-3):\penalty0 135--145,
  2008.

\bibitem[Svensson et~al.(2009)Svensson, Sandin, Cnattingius, Reilly, Pawitan,
  Hultman, and Lichtenstein]{18}
Anna~C Svensson, Sven Sandin, Sven Cnattingius, Marie Reilly, Yudi Pawitan,
  Christina~M Hultman, and Paul Lichtenstein.
\newblock Maternal effects for preterm birth: a genetic epidemiologic study of
  630,000 families.
\newblock \emph{American journal of epidemiology}, 170\penalty0 (11):\penalty0
  1365--1372, 2009.

\bibitem[Lawson et~al.(2013)Lawson, Cheverud, and Wolf]{19}
Heather~A Lawson, James~M Cheverud, and Jason~B Wolf.
\newblock Genomic imprinting and parent-of-origin effects on complex traits.
\newblock \emph{Nature Reviews Genetics}, 14\penalty0 (9):\penalty0 609--617,
  2013.

\bibitem[Bartolomei(2009)]{7}
Marisa~S Bartolomei.
\newblock Genomic imprinting: employing and avoiding epigenetic processes.
\newblock \emph{Genes \& development}, 23\penalty0 (18):\penalty0 2124--2133,
  2009.

\bibitem[Patten et~al.(2014)Patten, Ross, Curley, Queller, Bonduriansky, and
  Wolf]{8}
MM~Patten, L~Ross, JP~Curley, David~C Queller, R~Bonduriansky, and JB~Wolf.
\newblock The evolution of genomic imprinting: theories, predictions and
  empirical tests.
\newblock \emph{Heredity}, 113\penalty0 (2):\penalty0 119--128, 2014.

\bibitem[Zhou et~al.(2009)Zhou, Hu, Lin, and Fung]{41}
Ji-Yuan Zhou, Yue-Qing Hu, Shili Lin, and Wing~K Fung.
\newblock Detection of parent-of-origin effects based on complete and
  incomplete nuclear families with multiple affected children.
\newblock \emph{Human heredity}, 67\penalty0 (1):\penalty0 1--12, 2009.

\bibitem[Zhou et~al.(2010)Zhou, Ding, Fung, and Lin]{42}
Ji-Yuan Zhou, Jie Ding, Wing~K Fung, and Shili Lin.
\newblock Detection of parent-of-origin effects using general pedigree data.
\newblock \emph{Genetic epidemiology}, 34\penalty0 (2):\penalty0 151--158,
  2010.

\bibitem[Zhou et~al.(2012)Zhou, Mao, Li, Hu, Xia, and Fung]{43}
Ji-Yuan Zhou, Wei-Gao Mao, Dan-Ling Li, Yue-Qing Hu, Fan Xia, and Wing~Kam
  Fung.
\newblock A powerful parent-of-origin effects test for qualitative traits
  incorporating control children in nuclear families.
\newblock \emph{Journal of human genetics}, 57\penalty0 (8):\penalty0 500--507,
  2012.

\bibitem[Wolf and Wade(2009)]{32}
Jason~B Wolf and Michael~J Wade.
\newblock What are maternal effects (and what are they not)?
\newblock \emph{Philosophical Transactions of the Royal Society B: Biological
  Sciences}, 364\penalty0 (1520):\penalty0 1107--1115, 2009.

\bibitem[Lin(2013{\natexlab{a}})]{33}
Shili Lin.
\newblock Assessing the effects of imprinting and maternal genotypes on complex
  genetic traits.
\newblock \emph{Risk Assessment and Evaluation of Predictions}, pages 285--300,
  2013{\natexlab{a}}.

\bibitem[Lin(2013{\natexlab{b}})]{21}
Shili Lin.
\newblock Assessing the effects of imprinting and maternal genotypes on complex
  genetic traits.
\newblock \emph{Risk Assessment and Evaluation of Predictions}, pages 285--300,
  2013{\natexlab{b}}.

\bibitem[Yang and Lin(2013)]{20}
Jingyuan Yang and Shili Lin.
\newblock Robust partial likelihood approach for detecting imprinting and
  maternal effects using case-control families.
\newblock \emph{The Annals of Applied Statistics}, pages 249--268, 2013.

\bibitem[Weinberg et~al.(1998)Weinberg, Wilcox, and Lie]{22}
CR~Weinberg, AJ~Wilcox, and RT~Lie.
\newblock A log-linear approach to case-parent--triad data: assessing effects
  of disease genes that act either directly or through maternal effects and
  that may be subject to parental imprinting.
\newblock \emph{The American Journal of Human Genetics}, 62\penalty0
  (4):\penalty0 969--978, 1998.

\bibitem[Han et~al.(2013)Han, Hu, and Lin]{23}
Miao Han, Yue-Qing Hu, and Shili Lin.
\newblock Joint detection of association, imprinting and maternal effects using
  all children and their parents.
\newblock \emph{European Journal of Human Genetics}, 21\penalty0 (12):\penalty0
  1449--1456, 2013.

\bibitem[Zhang et~al.(2019)Zhang, Khalili, and Lin]{25}
Fangyuan Zhang, Abbas Khalili, and Shili Lin.
\newblock Imprinting and maternal effect detection using partial likelihood
  based on discordant sibpair data.
\newblock \emph{Statistica Sinica}, 29\penalty0 (4):\penalty0 1915--1937, 2019.

\bibitem[Aavani(2019)]{51}
Pooya Aavani.
\newblock \emph{Detecting Imprinting and Maternal Effects Using Monte Carlo
  Expectation Maximization Algorithm}.
\newblock PhD thesis, 2019.

\bibitem[Peyrard(2001)]{45}
Nathalie Peyrard.
\newblock \emph{Convergence of MCEM and related algorithms for hidden Markov
  random field}.
\newblock PhD thesis, INRIA, 2001.

\bibitem[Levine and Casella(2001)]{44}
Richard~A Levine and George Casella.
\newblock Implementations of the monte carlo em algorithm.
\newblock \emph{Journal of Computational and Graphical Statistics}, 10\penalty0
  (3):\penalty0 422--439, 2001.

\bibitem[Wei and Tanner(1990)]{46}
Greg~CG Wei and Martin~A Tanner.
\newblock A monte carlo implementation of the em algorithm and the poor man's
  data augmentation algorithms.
\newblock \emph{Journal of the American statistical Association}, 85\penalty0
  (411):\penalty0 699--704, 1990.

\bibitem[Boyles(1983)]{28}
Russell~A Boyles.
\newblock On the convergence of the em algorithm.
\newblock \emph{Journal of the Royal Statistical Society: Series B
  (Methodological)}, 45\penalty0 (1):\penalty0 47--50, 1983.

\bibitem[Xu and Jordan(1996)]{35}
Lei Xu and Michael~I Jordan.
\newblock On convergence properties of the em algorithm for gaussian mixtures.
\newblock \emph{Neural computation}, 8\penalty0 (1):\penalty0 129--151, 1996.

\bibitem[Murray(1977)]{36}
GD~Murray.
\newblock Discussion of the paper by professor dempster et al.
\newblock \emph{Journal of the Royal Statistical Society Series B},
  39:\penalty0 27--28, 1977.

\bibitem[Lange(1995)]{37}
Kenneth Lange.
\newblock A gradient algorithm locally equivalent to the em algorithm.
\newblock \emph{Journal of the Royal Statistical Society: Series B
  (Methodological)}, 57\penalty0 (2):\penalty0 425--437, 1995.

\bibitem[Neath(2013)]{34}
Ronald~C. Neath.
\newblock On convergence properties of the monte carlo em algorithm.
\newblock \emph{Advances in Modern Statistical Theory and Applications: A
  Festschrift in honor of Morris L. Eaton}, 10:\penalty0 43 -- 62, 2013.
\newblock URL \url{http://dx.doi.org/10.1214/12-IMSCOLL1003}.

\bibitem[Chan and Ledolter(1995)]{38}
KS~Chan and Johannes Ledolter.
\newblock Monte carlo em estimation for time series models involving counts.
\newblock \emph{Journal of the American Statistical Association}, 90\penalty0
  (429):\penalty0 242--252, 1995.

\bibitem[Brooks and Gelman(1998)]{64}
Stephen~P Brooks and Andrew Gelman.
\newblock General methods for monitoring convergence of iterative simulations.
\newblock \emph{Journal of computational and graphical statistics}, 7\penalty0
  (4):\penalty0 434--455, 1998.

\bibitem[Khor et~al.(2011)Khor, Davila, Breunis, Lee, Shimizu, Wright, Yeung,
  Tan, Sim, Wang, et~al.]{65}
Chiea~Chuen Khor, Sonia Davila, Willemijn~B Breunis, Yi-Ching Lee, Chisato
  Shimizu, Victoria~J Wright, Rae~SM Yeung, Dennis~EK Tan, Kar~Seng Sim,
  Jie~Jin Wang, et~al.
\newblock Genome-wide association study identifies fcgr2a as a susceptibility
  locus for kawasaki disease.
\newblock \emph{Nature genetics}, 43\penalty0 (12):\penalty0 1241--1246, 2011.

\bibitem[Chang et~al.(2016)Chang, Wang, Hsiung, He, Lin, Sheu, Chang,
  Quertermous, Chen, Rotter, et~al.]{53}
Tien-Jyun Chang, Wen-Chang Wang, Chao~A Hsiung, Chih-Tsueng He, Ming-Wei Lin,
  Wayne Huey-Herng Sheu, Yi-Cheng Chang, Tom Quertermous, Ida Chen, Jerome
  Rotter, et~al.
\newblock Genetic variation in the human sorbs1 gene is associated with blood
  pressure regulation and age at onset of hypertension: A sapphire cohort
  study.
\newblock \emph{Medicine}, 95\penalty0 (10), 2016.

\bibitem[Jung et~al.(2021)Jung, Elenbaas, Alisio, Santana, Young, Kang,
  Kachroo, Lavine, Razani, Mecham, et~al.]{54}
In-Hyuk Jung, Jared~S Elenbaas, Arturo Alisio, Katherine Santana, Erica~P
  Young, Chul~Joo Kang, Puja Kachroo, Kory~J Lavine, Babak Razani, Robert~P
  Mecham, et~al.
\newblock Svep1 is a human coronary artery disease locus that promotes
  atherosclerosis.
\newblock \emph{Science translational medicine}, 13\penalty0 (586):\penalty0
  eabe0357, 2021.

\bibitem[Isobe et~al.(2019)Isobe, Kataoka, Endo, Moriyama, Okazaki,
  Tsuchihashi, Katsumata, Yamamoto, Shirakawa, Yoshida, et~al.]{55}
Sarasa Isobe, Masaharu Kataoka, Jin Endo, Hidenori Moriyama, Shogo Okazaki,
  Kenji Tsuchihashi, Yoshinori Katsumata, Tsunehisa Yamamoto, Kohsuke
  Shirakawa, Naohiro Yoshida, et~al.
\newblock Endothelial--mesenchymal transition drives expression of cd44 variant
  and xct in pulmonary hypertension.
\newblock \emph{American journal of respiratory cell and molecular biology},
  61\penalty0 (3):\penalty0 367--379, 2019.

\bibitem[Jiang et~al.(2022)Jiang, Hei, Hao, Lin, Wang, Liu, Meng, and Guan]{56}
Yunfei Jiang, Bingchang Hei, Wenbo Hao, Shudong Lin, Yuanyuan Wang, Xuzhi Liu,
  Xianguo Meng, and Zhanjiang Guan.
\newblock Clinical value of lncrna sox2-ot in pulmonary arterial hypertension
  and its role in pulmonary artery smooth muscle cell proliferation, migration,
  apoptosis, and inflammatory.
\newblock \emph{Heart \& Lung}, 55:\penalty0 16--23, 2022.

\bibitem[Chan et~al.(2018)Chan, Huang, Sun, Lee, Lien, and Chang]{57}
Hao Chan, Hsuan-Shun Huang, Der-Shan Sun, Chung-Jen Lee, Te-Sheng Lien, and
  Hsin-Hou Chang.
\newblock Trpm8 and raas-mediated hypertension is critical for cold-induced
  immunosuppression in mice.
\newblock \emph{Oncotarget}, 9\penalty0 (16):\penalty0 12781, 2018.

\bibitem[Cai and Liu(2021)]{58}
Haomin Cai and Hongcheng Liu.
\newblock Immune infiltration landscape and immune-marker molecular typing of
  pulmonary fibrosis with pulmonary hypertension.
\newblock \emph{BMC Pulmonary Medicine}, 21\penalty0 (1):\penalty0 1--15, 2021.

\bibitem[Morrison et~al.(2002)Morrison, Bray, Folsom, and Boerwinkle]{59}
Alanna~C Morrison, Molly~S Bray, Aaron~R Folsom, and Eric Boerwinkle.
\newblock Add1 460w allele associated with cardiovascular disease in
  hypertensive individuals.
\newblock \emph{Hypertension}, 39\penalty0 (6):\penalty0 1053--1057, 2002.

\bibitem[Fenger et~al.(2011)Fenger, Linneberg, J{\o}rgensen, Madsbad, S{\o}bye,
  Eugen-Olsen, and Jeppesen]{60}
Mogens Fenger, Allan Linneberg, Torben J{\o}rgensen, Sten Madsbad, Karen
  S{\o}bye, Jesper Eugen-Olsen, and J{\o}rgen Jeppesen.
\newblock Genetics of the ceramide/sphingosine-1-phosphate rheostat in blood
  pressure regulation and hypertension.
\newblock \emph{BMC genetics}, 12\penalty0 (1):\penalty0 1--18, 2011.

\bibitem[Hadinnapola et~al.(2014)Hadinnapola, Southwood, Jenkins, Sheares,
  Toshner, and Papworth]{61}
C~Hadinnapola, M~Southwood, D~Jenkins, K~Sheares, M~Toshner, and J~Pepke-Zaba
  Papworth.
\newblock P16 a systematic characterisation of inflammation in chronic
  thromboembolic pulmonary hypertension.
\newblock \emph{Thorax}, 69\penalty0 (2):\penalty0 A1--A233, 2014.

\bibitem[Duarte et~al.(2012)Duarte, Zineh, Burkley, Gong, Langaee, Turner,
  Chapman, Boerwinkle, Gums, Cooper-DeHoff, et~al.]{62}
Julio~D Duarte, Issam Zineh, Ben Burkley, Yan Gong, Taimour~Y Langaee,
  Stephen~T Turner, Arlene~B Chapman, Eric Boerwinkle, John~G Gums, Rhonda~M
  Cooper-DeHoff, et~al.
\newblock Effects of genetic variation in h3k79 methylation regulatory genes on
  clinical blood pressure and blood pressure response to hydrochlorothiazide.
\newblock \emph{Journal of translational medicine}, 10\penalty0 (1):\penalty0
  1--9, 2012.

\end{thebibliography}
\appendix\section{Conditional distribution of subvector of a Dirichlet random variable}
Let, $Z = (Z_1,...,Z_k)$ follows a Dirichlet distribution with parameters $(\alpha_1,...,\alpha_k)$. \\
		Then, let $(1,...,k) = (A_1,A_2)$ and $Z = (Z_{A_1},Z_{A_2})$ \\
		By the definition of joint pdf of $(Z_1,...,Z_k)$;\\
		\begin{equation}
			f_{Z_1,...Z_k}(z_1,...,z_k) = \frac{\Gamma (\sum_{i=1}^{k}\alpha_i)}{\prod_{i=1}^{k} \Gamma(\alpha_i)} \prod_{i=1}^{k} z_i^{\alpha_i-1} , z_i \in (0,1), \sum_{i=1}^{k} z_i = 1
		\end{equation}
		
		Similarly, the joint pdf of $Z_{A_2}$ is;\\
		\begin{equation}
			f_{Z_{A_2}}(z_{A_2}) = \frac{\Gamma (\sum_{i=1}^{k}\alpha_i)}{\Gamma(\alpha_0) \prod_{i \in A_2} \Gamma(\alpha_i)} \prod_{i \in A_2} z_i^{\alpha_i - 1} \Bigg(1 - \sum_{i \in A_2} z_i\Bigg)^{\alpha_0 - 1}
		\end{equation}
		where, 
		$\alpha_0 = \sum_{i \in A_1} \alpha_i$. Then, the conditional pdf of $Z_{A_1}|Z_{A_2}$ is given by;\\
		\begin{equation}
			\begin{aligned}[b]
				f_{Z_{A_1}|Z_{A_2}}(z_{A_1}|z_{A_2}) & = \frac{\Gamma(\sum_{i \in A_1}\alpha_i)}{\prod_{i \in A_1}\Gamma(\alpha_i)} \prod_{i \in A_1} z_i^{\alpha_i-1} \Bigg(1 - \sum_{i \in A_2} z_i\Bigg)^{-(\alpha_0 - 1)}\\
				& = \frac{\Gamma(\sum_{i \in A_1}\alpha_i)}{\prod_{i \in A_1}\Gamma(\alpha_i)} \prod_{i \in A_1} \Bigg[z_i \Bigg(1 - \sum_{i \in A_2}z_i\Bigg)^{-1}\Bigg]^{\alpha_i - 1} \Bigg(1 - \sum_{i \in A_2} z_i\Bigg)^{-(m-1)}
			\end{aligned}
		\end{equation}
		
		When $m = $length$(A_1)$, $Z_{A_1}|Z_{A_2} = z_{A_2}$ has a scaled Dirichlet distribution;\\ 
		\begin{equation}
			\frac{1}{1 - \textbf{1}^T z_{A_2}} Z_{A_1}|Z_{A_2} = z_{A_2} \sim Dir(\alpha_{A_1})
		\end{equation}
		
		where \textbf{1} is the vector with length $k - m$ and all entries equal to 1.
		\clearpage
		\section{Calculation of Probabilities in Table 3.2}
		In Table 3.2, the formula to calculate the joint probability is as follows:\\
		\begin{align*}
			P(M  & = m,F = f, C_1 = c_1,C_2 = c_2,D_1 = 1,D_2 = 0)\\
			& = P(M = m,F = f).P(C_1 = c_1|M = m,F = f).P(C_2 = c_2|M = m,F = f)\\
			&\times P(D_1 = 1|M = m,F = f,C_1 = c_1).P(D_2 = 0|M = m,F = f,C_2 = c_2).
		\end{align*}
		
		For all types other than types 12, 13, 14, 18 and 19 (Table 3.2), if a child has one copy of the variant allele, the parental origin can be unambiguously identified, and hence the joint probability can be easily obtained by extracting the relevant factors from the relative risk models for disease prevalence:\\
		$P(D_1=1|M=m,F=f,C_1=c_1) = \delta R_{1}^{I(c_1=1)} R_{2}^{I(c_1=2)} R_{im}^{I(c_1 = 1_m)} S_{1}^{I(M=1)} S_{2}^{I(M=2)},$ \\
		$P(D_2=0|M=m,F=f,C_2=c_2) = 1 - \delta R_{1}^{I(c_2=1)} R_{2}^{I(c_2=2)} R_{im}^{I(c_2 = 1_m)} S_{1}^{I(M=1)} S_{2}^{I(M=2)},$\\
		For example, in the familial genotype combination $(m, f, c_1, c_2) = (2, 0, 1, 1)$, \\
		\begin{align*}
			P(M = 2,F = 0,C_1 = 1,C_2 & = 1,D_1 = 1,D_2 = 0)= P(M = 2,F = 0)\\
			&.P(C_1 = 1|M = 2,F = 0).P(C_2 = 1|M = 2,F = 0)\\
			&.P(D_1 = 1|M = 2,F = 0,C_1 = 1)\\
			&.P(D_2 = 0|M = 2,F = 0,C_2 = 1)\\
			& = \mu_{20} \delta r_1 s_2 r_{im} (1 - \delta r_1 s_2 r_{im}).\\
		\end{align*}
		For type 12, in which $(m, f, c_1, c_2) = (1, 1, 1, 0)$, as the variant allele carried by the affected child can be inherited either from the mother or the father with equal probabilities, the penetrance probability ends up being a equally weighted summation. \\
		That is, \\
		\begin{align*}
			P(D_1 = 1|M = 2,F = 0,C_1 = 1) & = \frac{1}{2} [\delta r_1 s_1 + \delta r_1 r_{im} s_1]\\
			& = \frac{1}{2} \delta r_1 s_1 (1 + r_{im}).\\
		\end{align*}
		Then the joint probability can be written as,\\
		\begin{align*}
			P(M = 1,F = 1,C_1 = 1& ,C_2 = 0,D_1 = 1,D_2 = 0) = P(M = 1,F = 1)\\
			&.P(C_1 = 1|M = 1,F = 1).P(C_2 = 0|M = 1,F = 1)\\
			&.P(D_1 = 1|M = 1,F = 1,C_1 = 1)\\
			&.P(D_2 = 1|M = 1,F = 1,C_2 = 0)\\
			& = \mu_{11} . \frac{1}{2} . \frac{1}{4} . \frac{1}{2} \delta r_1 s_1 (1 + r_{im}) . (1 - \delta s_1)\\
			& = \mu_{11} \frac{1}{16} \delta r_1 s_1 (1 + r_{im}) (1 - \delta s_1). \\
		\end{align*}
		For type 13, in which $(m, f, c_1, c_2) = (1, 1, 0, 1)$, the variant allele carried by the unaffected child can be inherited either from the mother or the father with equal probabilities and, as such, the penetrance probability ends up being the summation of two penetrance probabilities weighted equally. \\
		That is, \\
		\begin{align*}
			P(D_2 = 0|M = 1,F = 1,C_2 = 1) & = \frac{1}{2} [(1 - \delta r_1 s_1) + (1 - \delta r_1 s_1 r_{im})]\\
			& = \frac{1}{2} [2 - \delta r_1 s_1 (1 + r_{im})].
		\end{align*}
		Then, the joint probability can be written as,\\
		\begin{align*}
			P(M = 1,F = 1,C_1 = 0 & ,C_2 = 1,D_1 = 1,D_2 = 0) = P(M = 1,F = 1)\\
			&.P(C_1 = 0|M = 1,F = 1).P(C_2 = 1|M = 1,F = 1)\\
			&.P(D_1 = 1|M = 1,F = 1,C_1 = 0)\\
			&.P(D_2 = 0|M = 1,F = 1,C_2 = 1)\\
			& = \mu_{11} . \frac{1}{4} . \frac{1}{2} . \frac{1}{2} \delta s_1 . \frac{1}{2} [2 - \delta r_1 s_1 (1 + r_{im})]\\
			& = \mu_{11} \frac{1}{16} \delta s_1 [2 - \delta r_1 s_1 (1 + r_{im})] .\\
		\end{align*}
		For type 14, in which $(m, f, c_1, c_2) = (1, 1, 1, 1)$, the variant allele carried by the both affected and unaffected child can be inherited either from the mother or the father with equal probabilities and, as such, the penetrance probability ends up being a equally weighted summation. \\ 
		\begin{align*}
			P(M = 1,F = 1,C_1 = 1& ,C_2 = 1,D_1 = 1,D_2 = 0) = P(M = 1,F = 1)\\
			&.P(C_1 = 1|M = 1,F = 1).P(C_2 = 1|M = 1,F = 1)\\
			&.P(D_1 = 1|M = 1,F = 1,C_1 = 1)\\
			&.P(D_2 = 1|M = 1,F = 1,C_2 = 1)\\
			& = \mu_{11} . \frac{1}{2} . \frac{1}{2} . \frac{1}{2} \delta r_1 s_1 (1 + r_{im}) . \frac{1}{2} [2 - \delta r_1 s_1 (1 + r_{im})]\\
			& = \mu_{11} \frac{1}{16} \delta r_1 s_1 (1 + r_{im}) [2 - \delta r_1 s_1 (1 + r_{im})].\\
		\end{align*}
		We can follow similar procedures to derive the joint probabilities for other settings.
		\clearpage 
		\section{Results}
		\begin{figure}[ht!]
			\centering
			\includegraphics[width=\linewidth]{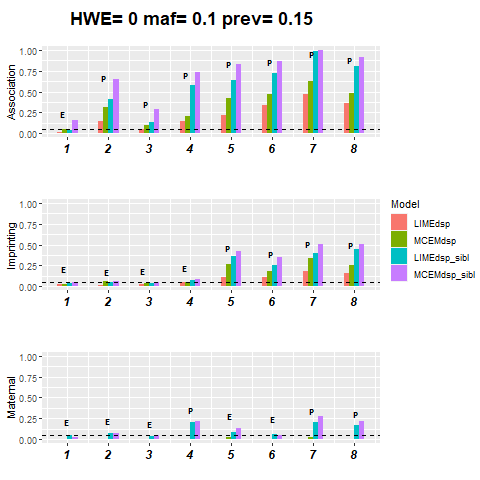}
			\caption{Bar charts for compare type I error and power of $MCEM_{DSP}$ and $LIME_{DSP}$ methods with and without additional siblings when HWE = 0, maf = 0.1 and prev = 0.15 with 100 families}
		\end{figure}
		\begin{figure}[ht!]
			\centering
			\includegraphics[width=\linewidth]{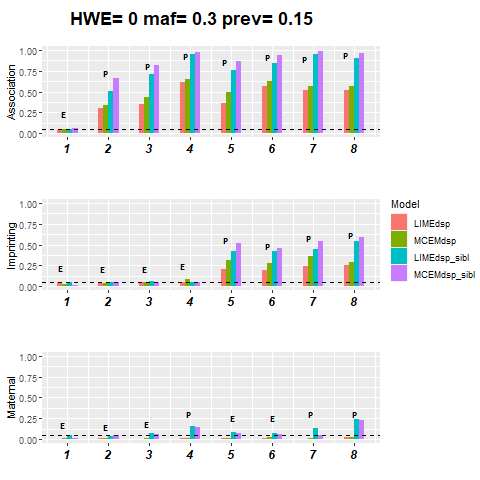}
			\caption{Bar charts for compare type I error and power of $MCEM_{DSP}$ and $LIME_{DSP}$ methods with and without additional siblings when HWE = 0, maf = 0.3 and prev = 0.15 with 100 families}
		\end{figure}
		
		\begin{figure}[ht!]
			\centering
			\includegraphics[width=\linewidth]{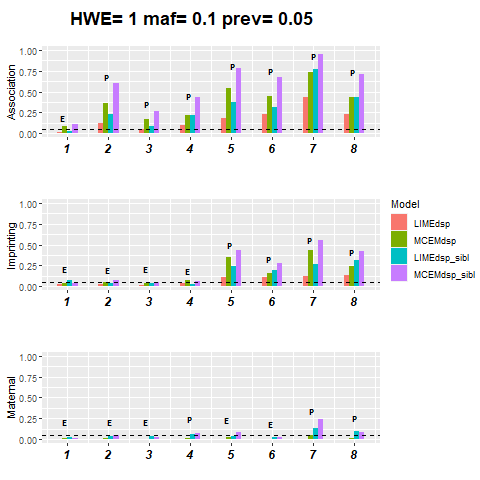}
			\caption{Bar charts for compare type I error and power of $MCEM_{DSP}$ and $LIME_{DSP}$ methods with and without additional siblings when HWE = 1, maf = 0.1 and prev = 0.05 with 100 families}
		\end{figure}
		\begin{figure}[ht!]
			\includegraphics[width=\linewidth]{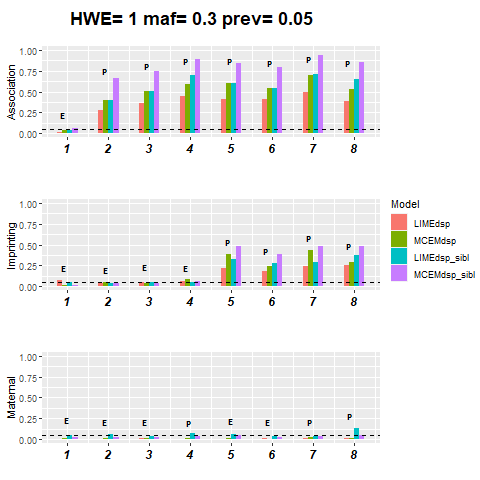}
			\caption{Bar charts for compare type I error and power of $MCEM_{DSP}$ and $LIME_{DSP}$ methods with and without additional siblings when HWE = 1, maf = 0.3 and prev = 0.05 with 100 families}
		\end{figure}
		\begin{figure}[ht!]
			\centering
			\includegraphics[width=\linewidth]{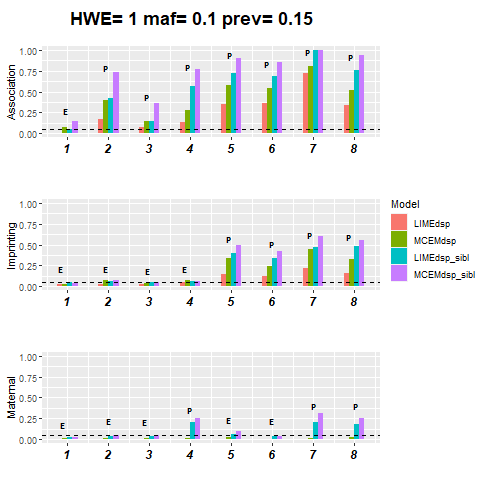}
			\caption{Bar charts for compare type I error and power of $MCEM_{DSP}$ and $LIME_{DSP}$ methods with and without additional siblings when HWE = 1, maf = 0.1 and prev = 0.15 with 100 families}
		\end{figure}
		\begin{figure}[ht!]
			\centering
			\includegraphics[width=\linewidth]{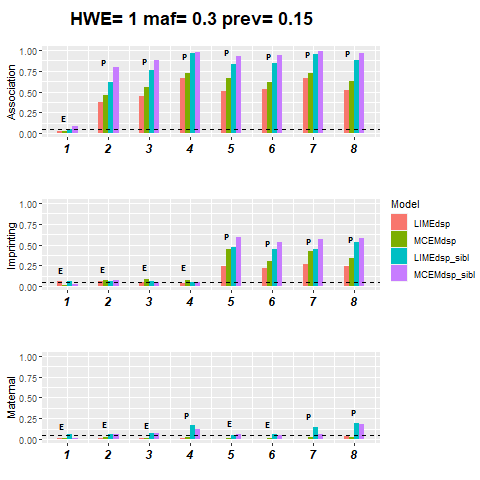}
			\caption{Bar charts for compare type I error and power of $MCEM_{DSP}$ and $LIME_{DSP}$ methods with and without additional siblings when HWE = 1, maf = 0.3 and prev = 0.15 with 100 families}
		\end{figure}
		
		\begin{figure}[ht!]
			\centering
			\includegraphics[width=\linewidth]{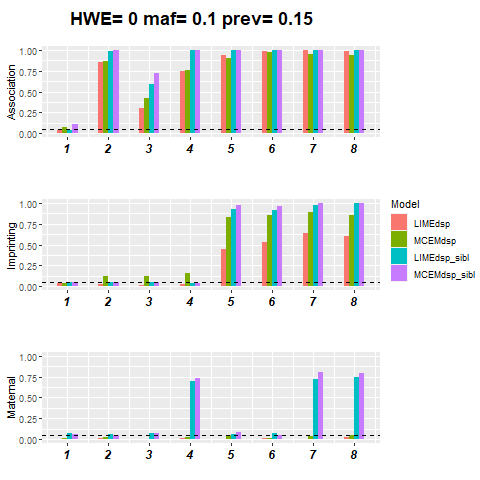}
			\caption{Bar charts for compare type I error and power of $MCEM_{DSP}$ and $LIME_{DSP}$ methods with and without additional siblings when HWE = 0, maf = 0.1 and prev = 0.15 with 500 families}
		\end{figure}
		\begin{figure}[ht!]
			\centering
			\includegraphics[width=\linewidth]{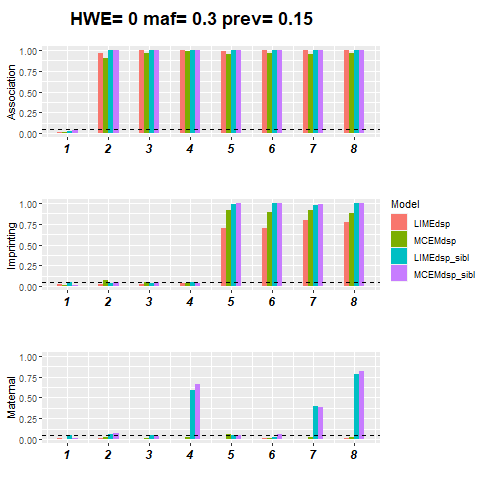}
			\caption{Bar charts for compare type I error and power of $MCEM_{DSP}$ and $LIME_{DSP}$ methods with and without additional siblings when HWE = 0, maf = 0.3 and prev = 0.15 with 500 families}
		\end{figure}
		
		\begin{figure}[ht!]
			\centering
			\includegraphics[width=\linewidth]{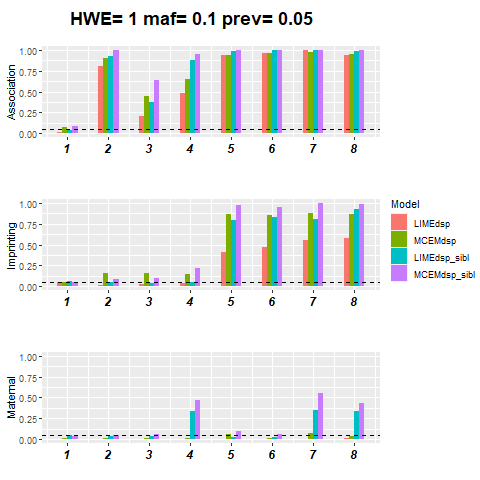}
			\caption{Bar charts for compare type I error and power of $MCEM_{DSP}$ and $LIME_{DSP}$ methods with and without additional siblings when HWE = 1, maf = 0.1 and prev = 0.05 with 500 families}
		\end{figure}
		\begin{figure}[ht!]
			\includegraphics[width=\linewidth]{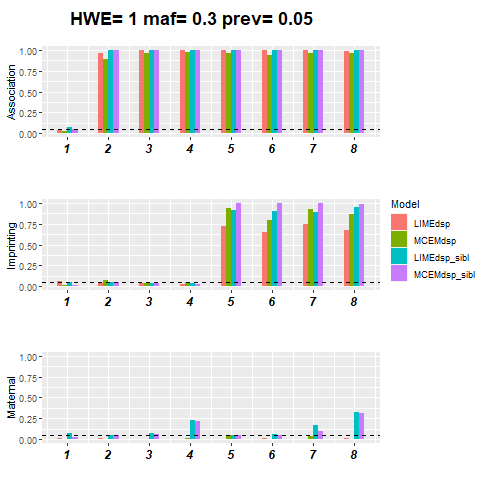}
			\caption{Bar charts for compare type I error and power of $MCEM_{DSP}$ and $LIME_{DSP}$ methods with and without additional siblings when HWE = 1, maf = 0.3 and prev = 0.05 with 500 families}
		\end{figure}
		\begin{figure}[ht!]
			\centering
			\includegraphics[width=\linewidth]{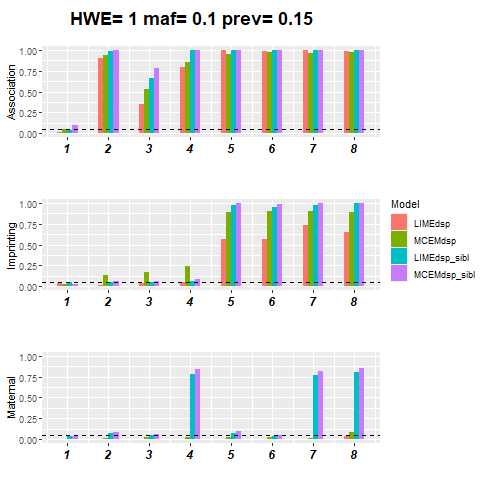}
			\caption{Bar charts for compare type I error and power of $MCEM_{DSP}$ and $LIME_{DSP}$ methods with and without additional siblings when HWE = 1, maf = 0.1 and prev = 0.15 with 500 families}
		\end{figure}
		\begin{figure}[ht!]
			\centering
			\includegraphics[width=\linewidth]{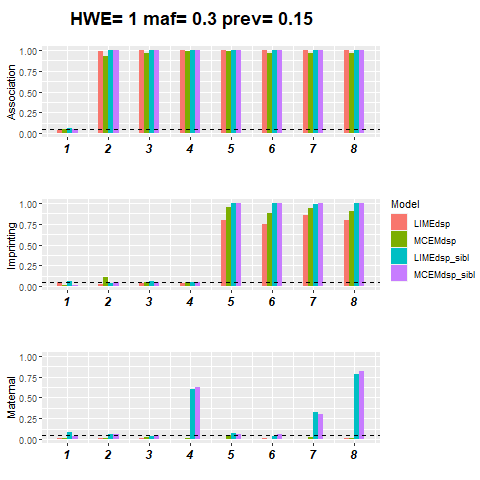}
			\caption{Bar charts for compare type I error and power of $MCEM_{DSP}$ and $LIME_{DSP}$ methods with and without additional siblings when HWE = 1, maf = 0.3 and prev = 0.15 with 500 families}
		\end{figure}
		
		\begin{figure}[ht!]
			\centering
			\includegraphics[width=0.9\textwidth]{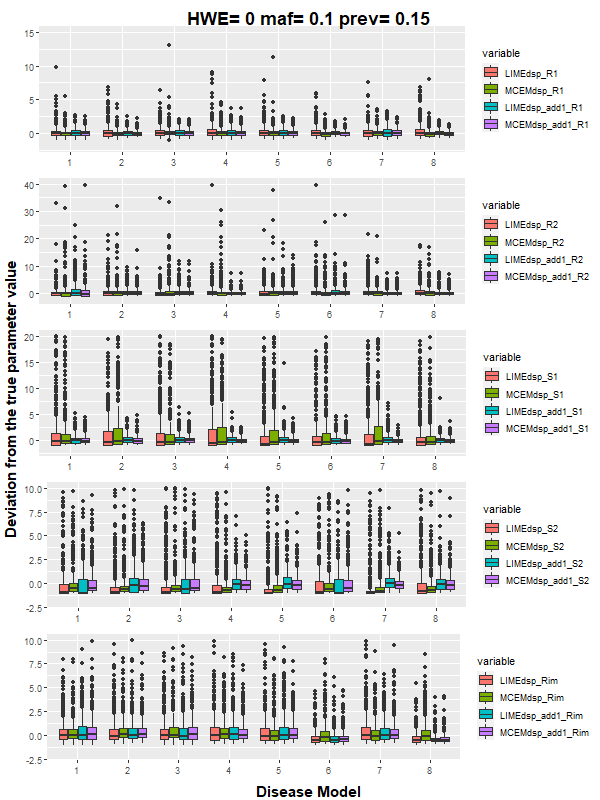}
			\caption{Box plots for biases of parameters from $MCEM_{DSP}$ and $LIME_{DSP}$ methods when HWE = 0, maf = 0.1 and PREV = 0.15 with 100 families}
		\end{figure}
		\begin{figure}[ht!]
			\centering
			\includegraphics[width=0.9\textwidth]{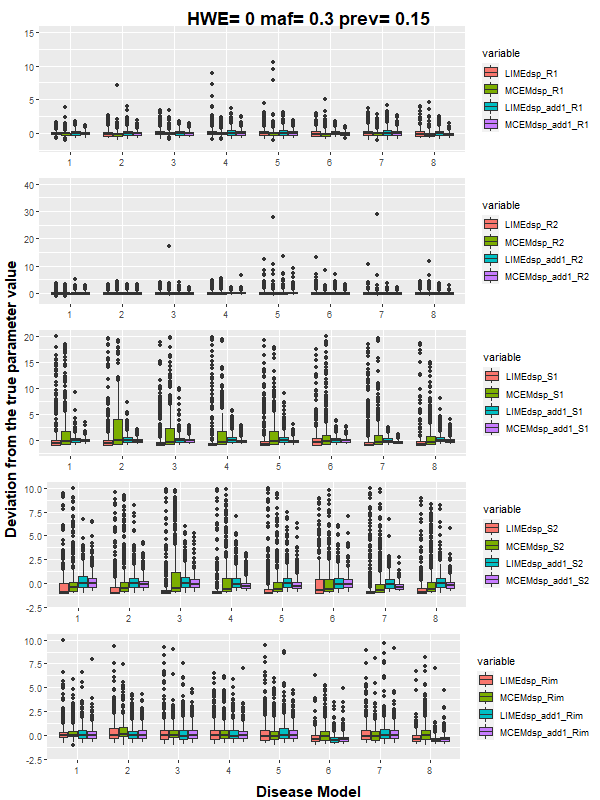}
			\caption{Box plots for biases of parameters from $MCEM_{DSP}$ and $LIME_{DSP}$ methods when HWE = 0, maf = 0.3 and PREV = 0.15 with 100 families}
		\end{figure}

		\begin{figure}[ht!]
			\centering
			\includegraphics[width=0.9\textwidth]{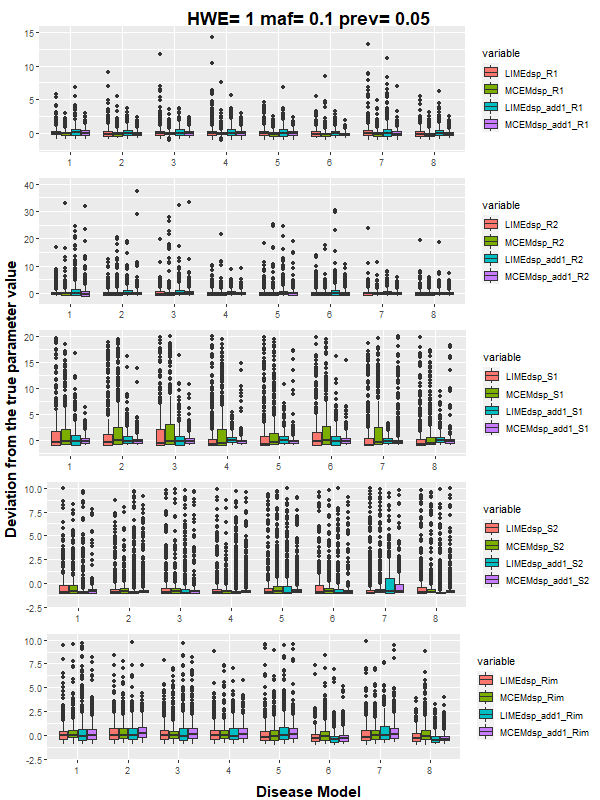}
			\caption{Box plots for biases of parameters from $MCEM_{DSP}$ and $LIME_{DSP}$ methods when HWE = 1, maf = 0.1 and PREV = 0.05 with 100 families}
		\end{figure}
		\begin{figure}[ht!]
			\centering
			\includegraphics[width=0.9\textwidth]{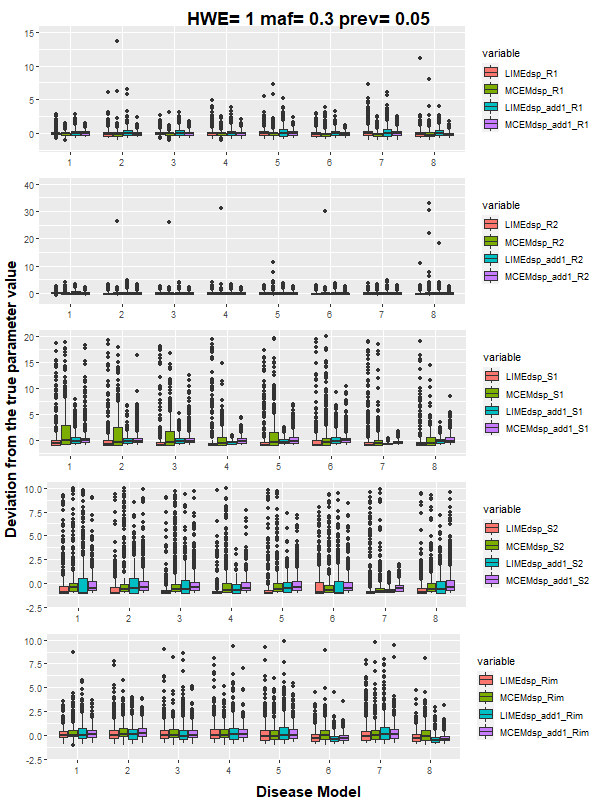}
			\caption{Box plots for biases of parameters from $MCEM_{DSP}$ and $LIME_{DSP}$ methods when HWE = 1, maf = 0.3 and PREV = 0.05 with 100 families}
		\end{figure}
		\begin{figure}[ht!]
			\centering
			\includegraphics[width=0.9\textwidth]{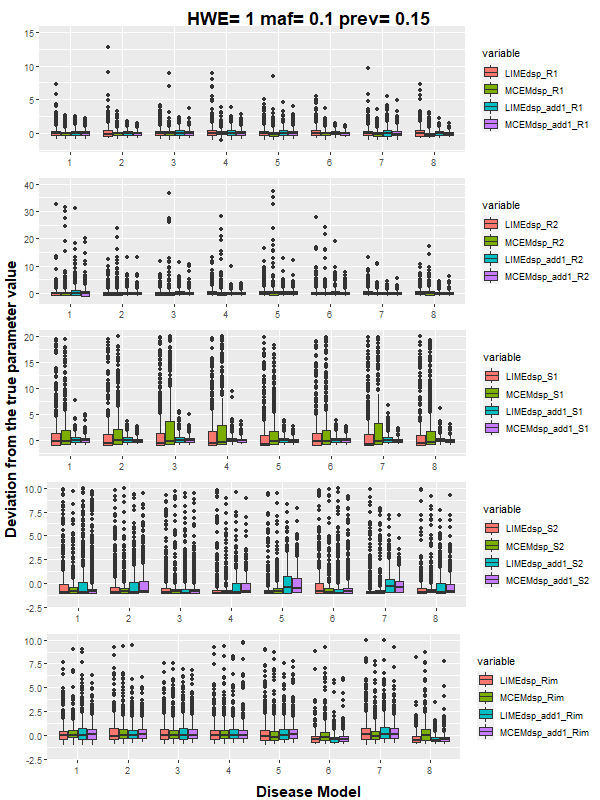}
			\caption{Box plots for biases of parameters from $MCEM_{DSP}$ and $LIME_{DSP}$ methods when HWE = 1, maf = 0.1 and PREV = 0.15 with 100 families}
		\end{figure}
		\begin{figure}[ht!]
			\centering
			\includegraphics[width=0.9\textwidth]{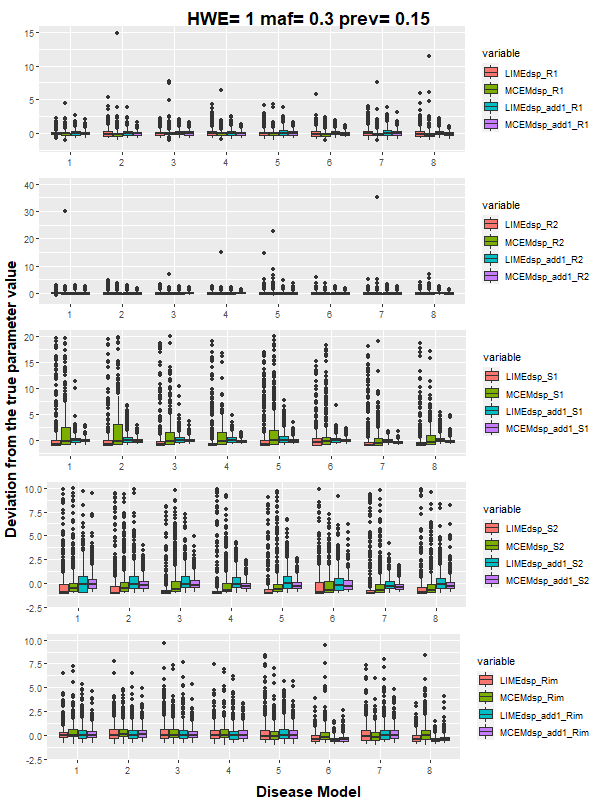}
			\caption{Box plots for biases of parameters from $MCEM_{DSP}$ and $LIME_{DSP}$ methods when HWE = 1, maf = 0.3 and PREV = 0.15 with 100 families}
		\end{figure}
		
		\begin{figure}[ht!]
			\centering
			\includegraphics[width=0.9\textwidth]{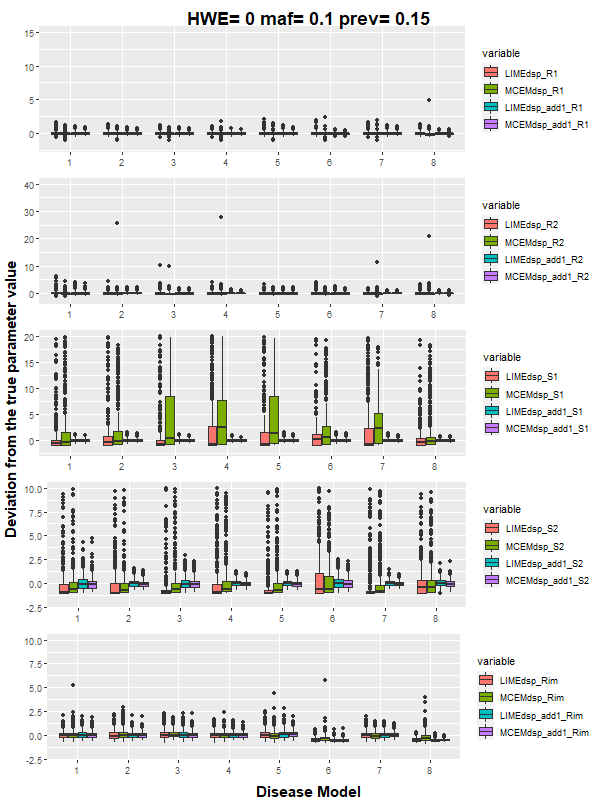}
			\caption{Box plots for biases of parameters from $MCEM_{DSP}$ and $LIME_{DSP}$ methods when HWE = 0, maf = 0.1 and PREV = 0.15 with 500 families}
		\end{figure}
		\begin{figure}[ht!]
			\centering
			\includegraphics[width=0.9\textwidth]{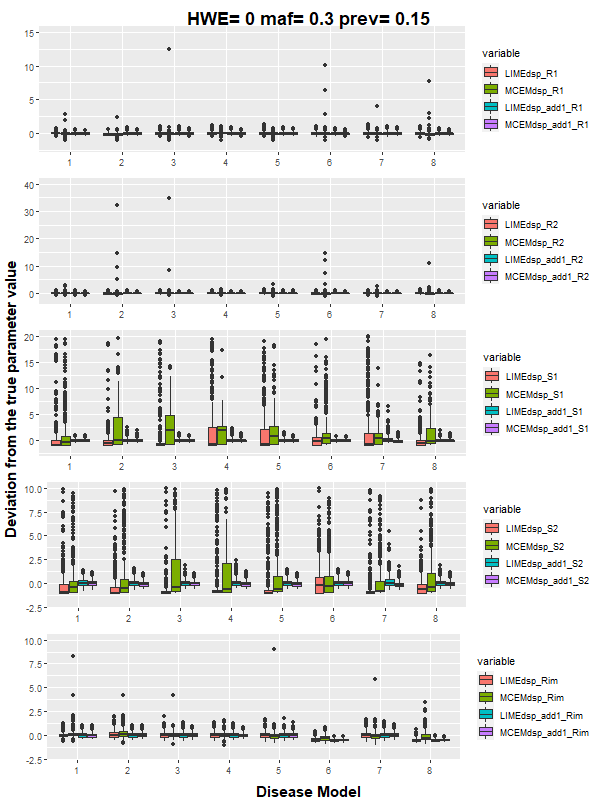}
			\caption{Box plots for biases of parameters from $MCEM_{DSP}$ and $LIME_{DSP}$ methods when HWE = 0, maf = 0.3 and PREV = 0.15 with 500 families}
		\end{figure}

		\begin{figure}[ht!]
			\centering
			\includegraphics[width=0.9\textwidth]{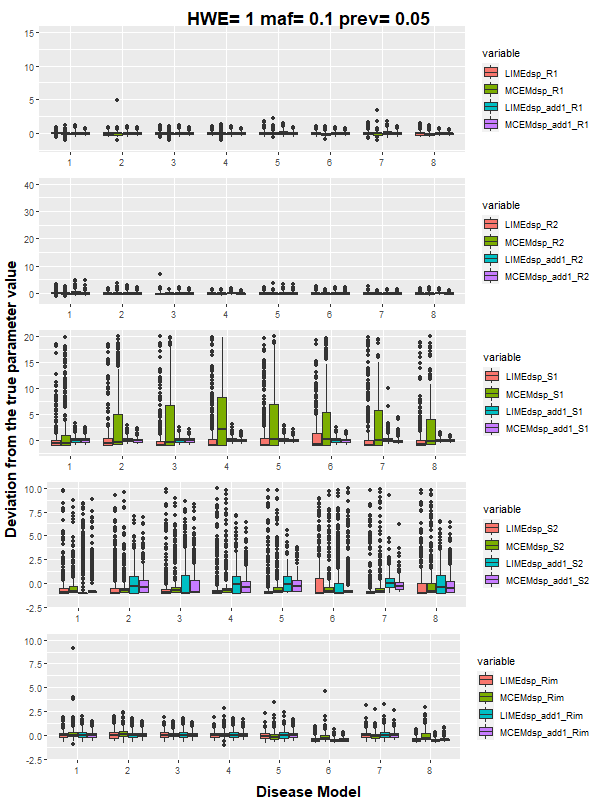}
			\caption{Box plots for biases of parameters from $MCEM_{DSP}$ and $LIME_{DSP}$ methods when HWE = 1, maf = 0.1 and PREV = 0.05 with 500 families}
		\end{figure}
		\begin{figure}[ht!]
			\centering
			\includegraphics[width=0.9\textwidth]{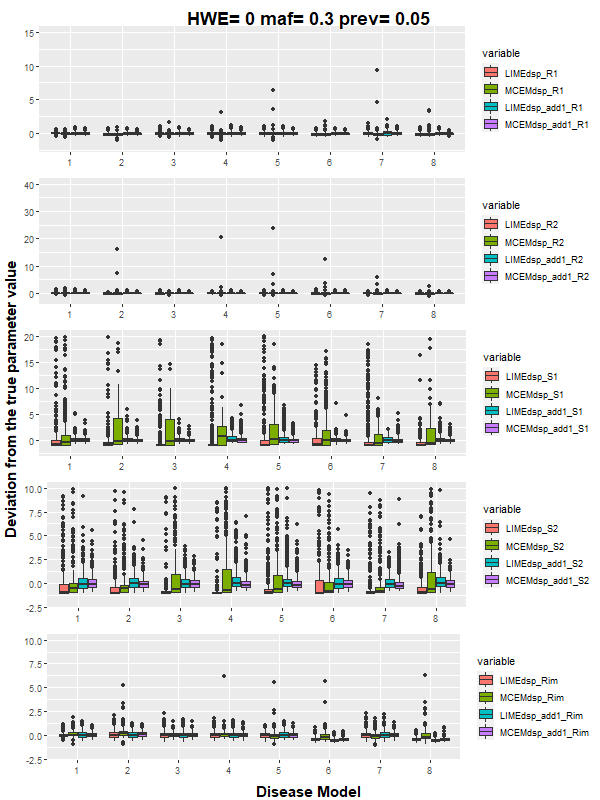}
			\caption{Box plots for biases of parameters from $MCEM_{DSP}$ and $LIME_{DSP}$ methods when HWE = 1, maf = 0.3 and PREV = 0.05 with 500 families}
		\end{figure}
		\begin{figure}[ht!]
			\centering
			\includegraphics[width=0.9\textwidth]{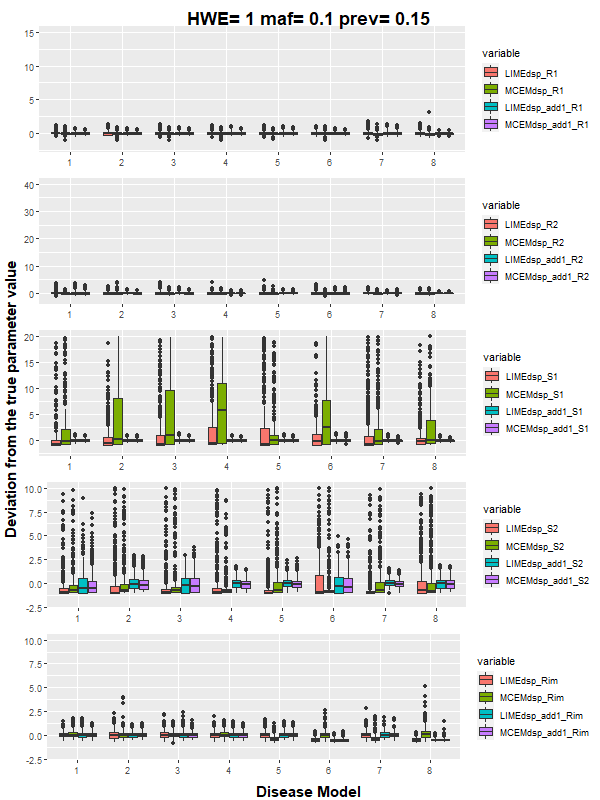}
			\caption{Box plots for biases of parameters from $MCEM_{DSP}$ and $LIME_{DSP}$ methods when HWE = 1, maf = 0.1 and PREV = 0.15 with 500 families}
		\end{figure}
		\begin{figure}[ht!]
			\centering
			\includegraphics[width=0.9\textwidth]{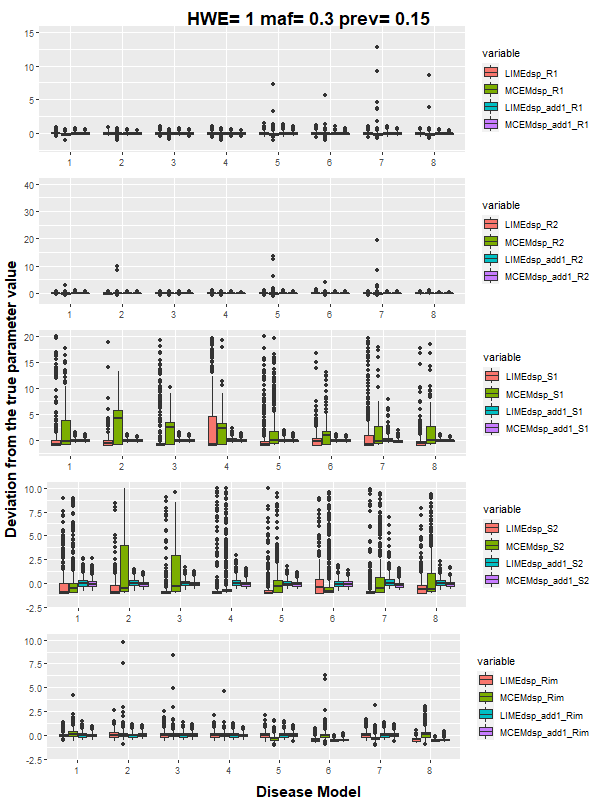}
			\caption{Box plots for biases of parameters from $MCEM_{DSP}$ and $LIME_{DSP}$ methods when HWE = 1, maf = 0.3 and PREV = 0.15 with 500 families}
		\end{figure}


\vspace{1cm} 
\section{Real data Analysis}
\begin{table}[!htb]
			\caption{Top 20 significant SNPs for association with Framingham Heart Study data using $MCEM_{DSP}$}
			\centering
			\begin{tabular}{| c| c| c| c| c |c| c |} 
				\hline
				Rank & SNP & Chr & Position(BP) & Gene & \multicolumn{2}{c}{$-log_{10}$P-value} \\
				\cline{6-7}
				& & & & &  $MCEM_{DSP}$ & $LIME_{DSP}$\\
				\hline
				1 & rs5760711 & 22 & 24894886 & SGSM1 & 15.9546 & 0.4676\\
				2 & rs12416299 & 10 & 95542074 & SORBS1 & 15.4775 & 0.2365\\
				3 & rs340719 & 2 & 16222132 & & 14.8754 & 0.4882\\
				4 & rs2973566 & 5 & 73852656 & ARHGEF28 & 14.8404 & 0.4387\\
				5 & rs11539522 & 15 & 49625193 & DTWD1 & 14.6122 & 0.4878\\
				6 & rs1146920 & 13 & 77760082 & SLAIN1 & 12.6497 & 0.4622\\
				7 & rs12117125 & 1 & 161522658 & FCGR2A & 12.3787 & 0.4241\\
				8 & rs2666954 & 6 & 380789 & & 12.0756 & 0.4285\\
				9 & rs6076157 & 20 & 23882207 & & 12.0674 & 0.4477\\
				10 & rs1327533 & 9 & 110368883 & SVEP1 & 12.0632 & 0.2198\\
				11 & rs12306837 & 12 & 4571302 & DYRK4 & 11.7532 & 0.4845\\
				12 & rs2281954 & 10 & 125796084 & UROS & 11.5686 & 0.4690\\
				13 & rs11033793 & 11 & 4769244 & MMP26 & 11.4601 & 0.1714\\
				14 & rs12345874 & 9 & 68776727 & PIP5K1B & 10.5720 & 0.4487\\
				15 & rs3829736 & 12 & 47982720 & COL2A1 & 10.2677 & 0.3841\\
				16 & rs9367018 & 6 & 12173899 & HIVEP1 & 9.6447 & 0.4840\\
				17 & rs12197569 & 6 & 2164221 & GMDS & 9.3021 & 0.4700\\
				18 & rs7124812 & 11 & 3439368 & LOC124902616 & 8.8759 & 0.3451\\
				19 & rs9382736 & 6 & 57436863 & PRIM2 & 8.8020 & 0.4176\\
				20 & rs16843643 & 2 & 241413353 & FARP2 & 8.6881 & 0.4622\\
				\hline	
			\end{tabular}
		\end{table}
		
		\begin{table}[!htb]
			\caption{Top 20 significant SNPs for imprinting with Framingham Heart Study data using $MCEM_{DSP}$}
			\centering
			\begin{tabular}{| c| c| c| c| c |c| c |} 
				\hline
				Rank & SNP & Chr & Position(BP) & Gene & \multicolumn{2}{c}{$-log_{10}$P-value} \\
				\cline{6-7}
				& & & & &  $MCEM_{DSP}$ & $LIME_{DSP}$\\
				\hline
				1 & rs2281954 & 10 & 125796084 & UROS & 12.7598 & 0.4690\\
				2 & rs7124812 & 11 & 3439368 & LOC124902616  & 8.3917 & 0.3451\\
				3 & rs7928271 & 11 & 126918754 & KIRREL3  & 5.9070 & 0.4653\\
				4 & rs6543298 & 2 & 98131244 & VWA3B & 4.9663 & 0.4565\\
				5 & rs2860161 & 7 & 142824583 &  & 4.8961 & 0.2271\\
				6 & rs6886844 & 5 & 105942041 &  & 4.8306 & 0.3260\\
				7 & rs6485742 & 11 & 12432529 & PARVA & 4.7765 & 0.3281\\
				8 & rs2059928 & 15 & 22152346 &  & 4.7728 & 0.3815\\
				9 & rs3734945 & 7 & 138621037 & SVOPL & 4.7018 & 0.3498\\
				10 & rs1729086 & 2 & 231343344 & ARMC9 & 4.0955 & 0.1001\\
				11 & rs2282537 & 11 & 120317262 & POU2F3 & 3.9273 & 0.1291\\
				12 & rs1106337 & 1 & 167166482 & LOC105371601 & 3.7137 & 0.0351\\
				13 & rs353637 & 11 & 35163005 & CD44 & 3.6998 & 0.1287\\
				14 & rs13209741 & 6 & 1528900 & LOC102723944 & 3.6427 & 0.1949\\
				15 & rs16832560 & 3 & 181564785 & SOX2-OT & 3.6316 & 0.1060\\
				16 & rs818055 & 9 & 131092123 & LAMC3 & 3.6123 & 0.2010\\
				17 & rs10018350 & 4 & 6320374 & PPP2R2C & 3.5524 & 0.1383\\
				18 & rs949292 & 18 & 60501574 & & 3.5417 & 0.3216\\
				19 & rs6878253 & 5 & 132277600 & LOC124901063 & 3.5211 & 0.0858\\
				20 & rs7593557 & 2 & 233955144 & TRPM8 & 3.5197 & 0476\\
				\hline	
			\end{tabular}
		\end{table}

		\begin{table}[!htb]
			\caption{Top 20 significant SNPs for maternal with Framingham Heart Study data using $MCEM_{DSP}$}
			\centering
			\begin{tabular}{| c| c| c| c| c |c| c |} 
				\hline
				Rank & SNP & Chr & Position(BP) & Gene & \multicolumn{2}{c}{$-log_{10}$P-value} \\
				\cline{6-7}
				& & & & &  $MCEM_{DSP}$ & $LIME_{DSP}$\\
				\hline
				1 & rs9294284 & 6 & 83954925 & CYB5R4 & 279.1653 & 0.4817 \\
				2 & rs4660212 & 1 & 42030728 & HIVEP3 &	6.6689 & 0.4603 \\
				3 & rs6485742 & 11 & 12432529 & PARVA & 6.4115	& 0.3281\\
				4 & rs7835497 & 8 & 39765231 & ADAM2 & 5.61404 & 0.1476\\
				5 & rs8130490 & 21 & 18355236 & TMPRSS15 & 5.5795 & 0.4691 \\
				6 & rs5022654 & 9 & 4766378	&  & 4.8936 & 0.0844 \\
				7 & rs12306837 & 12 & 4571302 & DYRK4 & 4.6017	& 0.4845 \\
				8 & rs7928271 & 11 & 126918754 & KIRREL3 & 4.5619 & 0.4643 \\
				9 & rs3751020 & 11 & 119608340 &  &	4.5395	& 0.3099\\
				10 & rs4961 & 4 & 2904980 & ADD1 & 4.4915 & 0.1957\\
				11 & rs12197569 & 6 & 2164221 & GMDS &	4.3880	& 0.4700 \\
				12 & rs12703608 & 7 & 144547846 & TPK1 & 4.3071 & 0.0997 \\
				13 & rs3811515 & 2 & 228018313 & SPHKAP & 4.2972 & 0.2384 \\
				14 & rs4814626 & 20 & 17580396 & DSTN &	4.2546	& 0.1779 \\
				15 & rs17594685 & 13 & 41679215 & VWA8 & 4.1531 & 0.1108 \\
				16 & rs7129737 & 11 & 124888741 & ROBO4 & 4.1144 & 0.1369 \\
				17 & rs12986413 & 19 & 2170955 & DOT1L & 4.0802 & 0.4825 \\
				18 & rs10807372 & 6 & 47681838 & ADGRF2 & 4.066 & 0.3564 \\
				19 & rs10188832 & 2 & 16621901 & CYRIA & 4.0573 & 0.2162 \\
				20 & rs11595441 & 10 & 15010629 &  & 4.0397 & 0.4536\\
				
				\hline	
			\end{tabular}
		\end{table}	
\end{document}